\documentclass[a4paper,onecolumn,longbibliography,accepted=2026-04-05]{quantumarticle}
\pdfoutput=1

\usepackage{amsmath,braket,amssymb}
\usepackage[final]{graphicx}
\usepackage{subfigure}
\usepackage{xcolor}
\usepackage[english]{babel}
\usepackage[utf8]{inputenc}
\usepackage{color}
\usepackage[sort&compress,numbers]{natbib}
\usepackage{hyperref}

\usepackage{algorithmicx}
\usepackage{algpseudocode}
\usepackage{amsmath}
\usepackage{comment}
\usepackage{amsthm}
\usepackage{bbold}
\usepackage{graphicx}
\usepackage{geometry}
\usepackage{braket}
\usepackage{afterpage}

\usepackage{booktabs}

\newcounter{algctr}
\renewcommand{\thealgctr}{\arabic{algctr}}

\newenvironment{Algorithm}[1]{%
  \refstepcounter{algctr}%
  \par\medskip
  \noindent\rule{\linewidth}{0.4pt}\par        % top rule
  \noindent\textbf{Algorithm~\thealgctr. #1}\par
  \noindent\rule{\linewidth}{0.4pt}\par        % rule directly under the title
  \vspace{0.em}%
  \begin{algorithmic}%
}{%
  \end{algorithmic}%
  \vspace{0.em}%
  \noindent\rule{\linewidth}{0.4pt}\par        % bottom rule
  \medskip
}

\usepackage{rotating}
\usepackage{multirow}

\usepackage{tcolorbox}
\usepackage{xcolor}
\tcbuselibrary{listings,skins} % <-- load both
\usepackage{sourcecodepro} % monospaced font
\tcbuselibrary{breakable, listings}

\definecolor{darkgreen}{rgb}{0,0.60,.2}
\definecolor{darkblue}{rgb}{0.1,0.3,1}

\usepackage{hyperref}
\hypersetup{
    colorlinks=true,
    linkcolor=blue,
    filecolor=magenta,      
    urlcolor=blue,
    citecolor=blue,
    pdftitle={Computing magic with HadaMAG}
    }
\usepackage{orcidlink}
\DeclareMathOperator{\Tr}{Tr}

\definecolor{purple}{HTML}{C792EA}
\definecolor{lightblue}{HTML}{82AAFF}
\definecolor{orange}{HTML}{F78C6C} % This orange will be used for true/false
\definecolor{yellow}{HTML}{FFCB6B}
\definecolor{gray}{HTML}{546E7A}
\definecolor{black}{HTML}{000000}
\definecolor{nightbg}{HTML}{2C2C3D}
\definecolor{framegray}{HTML}{3C3C3C}
\definecolor{creamyGold}{HTML}{FFE082}  % soft, warm yellow
\definecolor{juliaCyan}{HTML}{00DCDC}

\lstdefinelanguage{JuliaCustom}{
    keywords={function, return, end, where, for, if, else, elseif,break, using,begin},
    morekeywords=[2]{true, false}, % Adding true and false as a second set of keywords
    sensitive=true,
    morecomment=[l]{\#},
    morestring=[b]",
    alsoletter={!},
}

\newtcblisting{juliabox}{
    enhanced,
    colback=nightbg!95,
    colframe=framegray,
    boxrule=0.8pt,
    arc=1.5mm,
    outer arc=1.5mm,
    top=0.mm,    % Reduced internal space at the top
    bottom=0.mm, % Reduced internal space at the bottom
    left=2mm,
    right=2mm,
    listing only,
    listing engine=listings,
    listing options={language=JuliaCustom},
    before skip=\baselineskip,
    after skip=\baselineskip
}

\begin{document}
\title{Computing quantum magic of state vectors}

\author{Piotr Sierant \orcidlink{0000-0001-9219-7274}}
%\email{piotr.sierant@bsc.es}
\affiliation{Barcelona Supercomputing Center, Barcelona 08034, Spain}

\author{Jofre Vallès-Muns \orcidlink{0009-0000-0139-9744}}
\affiliation{Barcelona Supercomputing Center, Barcelona 08034, Spain}

\author{Artur Garcia-Saez  \orcidlink{0000-0003-3561-0223} }
\affiliation{Barcelona Supercomputing Center, Barcelona 08034, Spain}
\affiliation{Qilimanjaro Quantum Tech, 08019 Barcelona, Spain}

%\date{\today}

\begin{abstract}
Non-stabilizerness, also known as ``magic,'' quantifies how far a quantum state departs from the stabilizer set. It is a central resource behind quantum advantage and a useful probe of the complexity of quantum many-body states. Yet standard magic quantifiers, such as the stabilizer R\'enyi entropy (SRE) for qubits and the mana for qutrits, are costly to evaluate numerically, with the computational complexity growing rapidly with the number \(N\) of qudits. Here we introduce efficient, numerically exact algorithms that exploit the fast Hadamard transform to compute the SRE for qubits (\(d=2\)) and the mana for qutrits (\(d=3\)) for pure states given as state vectors. Our methods compute SRE and mana at cost \(O(N d^{2N})\), providing an \emph{exponential} improvement over the naive \(O(d^{3N})\) scaling, with substantial parallelism and straightforward GPU acceleration. We further show how to combine the fast Hadamard transform with Monte Carlo sampling to estimate the SRE of state vectors, and we extend the approach to compute the mana of mixed states. All algorithms are implemented in the open-source Julia package \href{https://github.com/bsc-quantic/HadaMAG.jl/}{\text{HadaMAG.jl}}, which provides a high-performance toolbox for computing SRE and mana with built-in support for multithreading, MPI-based distributed parallelism, and GPU acceleration. The package, together with the methods developed in this work, offers a practical route to large-scale numerical studies of magic in quantum many-body systems.
\end{abstract}

\maketitle

\section{Introduction} 
A quantum state of \(N\) qudits is represented by a vector in a Hilbert space of dimension \(\mathcal{N}=d^N\). This \textit{exponential} growth suggests that sufficiently large quantum devices may become intractable to simulate classically, enabling quantum computational advantage~\cite{preskill12quantum, daley22practical}. At the same time, many physically relevant states possess structure that can be leveraged for compact classical descriptions. For example, weakly entangled states admit efficient tensor-network representations~\cite{Vidal03a, Orus14, Ors2019, Ran20, Banuls23}, which have been central to our understanding of equilibrium many-body physics. In particular, ground states of one-dimensional local gapped Hamiltonians are well captured by matrix product states~\cite{Verstraete08, Schuch08, Schollw_ck_2011}. Beyond tensor networks, other families of many-body states are also classically tractable, including fermionic Gaussian states~\cite{Valiant01, Terhal02, Bravyi05flo} and stabilizer states~\cite{Gottesman1998, aaronson2004improvedsimulationof}. Although such states can exhibit extensive entanglement, they remain efficiently simulable with resources that scale \textit{polynomially} in system size $N$. This motivates quantitative notions of fermionic non-Gaussianity~\cite{Sierant25fermionicmagic} and non-stabilizerness~\cite{liu2022manybody}, which characterize how far a state lies from efficiently simulable classes such as Gaussian or stabilizer states.

Measures of non-stabilizerness are monotonic under stabilizer protocols~\cite{Veitch2014theresourcetheory, Heimendahl22, Haug23monotones}, which include Clifford operations~\cite{nielsen00}, but can increase under non-stabilizer gates such as the \(T\) gate~\cite{Bravyi05}. Together with Clifford gates, the \(T\) gate completes a universal gate set~\cite{Barenco95}. Several measures of non-stabilizerness have been proposed, including stabilizer rank~\cite{Bravyi16, Bravyi16improved}, stabilizer fidelity~\cite{Bravyi19simulatio}, and robustness of magic~\cite{Howard17, Heinrich19robustness, Sarkar20rom}. However, evaluating these measures typically requires costly optimization procedures that become prohibitive already for systems with fewer than \(N=10\) qubits~\cite{Hamaguchi24handbook}. 

In contrast, the recently introduced \emph{stabilizer R\'enyi entropy} (SRE)~\cite{leone2022stabilizerrenyientropy} avoids optimization while retaining the key properties expected of a qubit (\(d=2\)) non-stabilizerness measure~\cite{leone2024stabilizer}. Its introduction has sparked extensive efforts to characterize non-stabilizerness across a broad range of many-body settings, including quantum phases and phase transitions~\cite{haug2023quantifying, tarabunga23gauge, Odavic23, Passarelli24, Tarabunga24rk, Falcao25, Jasser2025, Bera2025SYK, ding2025sre, hoshino2025conformal, hoshino2025codes}, nonequilibrium dynamics~\cite{rattacaso2023stabilizer, Turkeshi25spectrum, turkeshi2024magic, Tirrito24anti, Odavic25, haug2024saturation, kos2024exact, tarabunga24transition, Santra25complexity, Passarelli25boundary, Hou2025highway, zhang2024quantumma, magni2025anticoncentrationcliffordcircuitsbeyond, magni2025quantumcomplexitychaosmanyqudit, zhang2025designsmagicaugmentedcliffordcircuits, szombathy25}, and experimental measurements~\cite{Oliviero2022measuring, Haug23scalable, Haug24efficient}. 

For qutrits (\(d=3\)), the structure of the Pauli group~\cite{Sarkar2024qudit} enables the definition of a discrete Wigner function~\cite{gross2006hudsonstheoremfor, Gross2007}, whose negativity yields a non-stabilizerness measure known as \emph{mana}~\cite{Veitch12, Veitch2014theresourcetheory, Pashayan15}. Like the SRE, mana avoids costly minimization, which has enabled studies in many-body ground states~\cite{White21, tarabunga2023critical} as well as in out-of-equilibrium settings~\cite{white2020manahaarrandomstates, Goto22, Sewell22, Koukoulekidis22, Basu2024, Ahmadi24, turkeshi2024magic, zhang2024quantumma, Nystrom25}. More broadly, investigations of non-stabilizerness extend beyond quantum computation and condensed matter physics, with applications in nuclear physics~\cite{Robin24nuclear, Brokemeier24quantum}, particle physics~\cite{White24quarks, Chernyshev25}, quantum electrodynamics~\cite{liu2025ed}, quantum chemistry~\cite{Gu2024zerofinite}, and the AdS/CFT correspondence~\cite{Goto22, cao2025gravitational}.

The ability to compute measures of non-stabilizerness is central to understanding magic-state resources in many-body systems. Although the SRE and mana avoid explicit minimization, their evaluation remains challenging because it requires computing \(\mathcal{N}^2=d^{2N}\) expectation values of highly nonlocal observables. For quantum states given as state vectors, a direct evaluation of these \(d^{2N}\) quantities~\cite{Ballar2025} is feasible on present-day hardware only up to \(N\approx 15\) qubits~\cite{Brokemeier24quantum, Bera2025SYK, Trigueros2025, Odavic23}. For weakly entangled states, which are well approximated by tensor networks, several approaches have been developed to compute the SRE~\cite{haug2023quantifying, lami2023perfect, tarabunga23gauge, tarabunga2024mps, liu2025translational} and mana~\cite{tarabunga2023critical}, enabling studies of systems comprising hundreds of qudits. 

However, many physically relevant quantum states are not weakly entangled. Examples include nonequilibrium dynamics~\cite{Luitz17, Sierant25mbl}, highly excited eigenstates of many-body Hamiltonians~\cite{Sierant20polfed, Morningstar22}, and systems with long-range interactions~\cite{Luitz19, Defenu24}. In such settings, numerical methods can still produce state vectors for systems with more than \(N=30\) qubits~\cite{Brenes17}, far beyond the regime where direct SRE evaluation is practical~\cite{Ballar2025} and, for highly entangled states, beyond the reach of tensor-network-based methods of non-stabilizerness calculation.

This work fills this gap by introducing efficient algorithms to compute the SRE and mana for quantum states represented as state vectors. Leveraging the fast Hadamard transform~\cite{Fino76}, we develop numerically exact procedures for evaluating SRE and mana that achieve an exponential speedup in the system size \(N\) compared to direct numerical methods~\cite{Ballar2025}, without additional memory overhead. This enables numerically exact computations of these non-stabilizerness measures for systems with up to \(N=25\) qubits and \(N=15\) qutrits. We further use the Hadamard transform to devise a sampling-based estimator for the SRE, extending the range of system sizes for which SRE can be computed from state vectors. Finally, we introduce a fast-Hadamard-transform-based method to compute the mana of mixed states.

This manuscript is structured as follows. In Sec.~\ref{sec:nonstab} we
introduce the non-stabilizerness measures of interest: the SRE and mana. Sec.~\ref{sec:exact} contains our main
contributions and presents
(i) in Sec.~\ref{subsec:effSRE}, a numerically exact algorithm for SRE;
(ii) in Sec.~\ref{subsec:sre_mc}, a sampling algorithm for SRE;
(iii) in Sec.~\ref{subsec:manafast}, a numerically exact algorithm for the
mana of pure states; and
(iv) in Sec.~\ref{subsec:mana_mixed_fast_sub}, an algorithm for the mana of mixed
states.
All these algorithms are implemented in the open-source package
\href{https://github.com/bsc-quantic/HadaMAG.jl/}{HadaMAG.jl}.
In Sec.~\ref{subsec:uses} we demonstrate the use of this package and
benchmark our methods on states generated by random quantum circuits.
Concluding remarks and an outlook are given in Sec.~\ref{sec:conclusion}.

\section{Non-stabilizerness measures}
\label{sec:nonstab}

We consider a system of \(N\) qubits (\(d = 2\)), and denote by \(I\), \(X\), \(Y\), and \(Z\) the identity and the Pauli \(X\), \(Y\), and \(Z\) operators.
Pauli strings are operators of the form
\(P = P_1 \otimes P_2 \otimes \cdots \otimes P_N\),
where each \(P_j \in \{I, X, Y, Z\}\).
The set of all \(N\)-qubit Pauli strings (modulo phases \(\pm 1, \pm i\)) will be denoted by \(\mathcal{P}_N\).
An element \(P\in\mathcal{P}_N\) can be parametrized by two binary strings
\(\mathbf a = (a_1,\dots,a_N) \in \mathbb{Z}_2^N \) and \(\mathbf b = (b_1,\dots,b_N) \in \mathbb{Z}_2^N \),
with \(a_j,b_j\in\{0,1\}\), as
\begin{equation}
  P_{\mathbf a,\mathbf b}
  =
\prod_{j=1}^N X_j^{a_j}\prod_{j=1}^N Z_j^{b_j},
\end{equation}
up to an overall phase.
The Clifford group \(\mathcal{C}_{N,2}\) is the subgroup of the unitary group
that maps Pauli strings to Pauli strings (up to phases \(\pm 1, \pm i\)).
Stabilizer states, denoted by \(\ket{\sigma}\), are the states obtained by
the action of \(\mathcal{C}_{N,2}\) on the initial  state
\(\ket{0}^{\otimes N}\).

The stabilizer Rényi entropy (SRE), introduced in Ref.~\cite{leone2022stabilizerrenyientropy}, is defined for a pure state $\ket{\psi}$ of $N$ qubits as
\begin{equation}
M_{q}(\ket{\psi}) = \frac{1}{1 - q} \log_2\left[ \,\sum_{P \in \mathcal{P}_N} \frac{\langle \psi | P | \psi \rangle^{2q}}{2^N} \right]
=\frac{1}{1 - q} \log_2\left[ \,\sum_{ \mathbf a, \mathbf b \in \mathbb{Z}_2^N} \frac{\langle \psi | P_{\mathbf{a},\mathbf{b}} | \psi \rangle^{2q}}{2^N} \right]
, \label{eq:sre}
\end{equation}
where $q > 0$ is the Rényi index, and the sum extends over all $4^N$ Pauli strings $ P_{\mathbf{a},\mathbf{b}}$. The quantity $M_1$ is defined via the limit $q \to 1$ in \eqref{eq:sre}.

The SRE satisfies the key properties expected of a non-stabilizerness measure for many-body systems:
\begin{itemize}
\item faithfulness: $M_q(\ket{\psi}) \ge 0$, with equality if and only if $\ket{\psi}$ is a stabilizer state\;
\item Clifford invariance: $M_q(C\ket{\psi}) = Mq(\ket{\psi})$ for any Clifford unitary $C \in \mathcal{C}_{N,2}$\;
% , i.e., any unitary that maps Pauli strings $P$ to Pauli strings $P' = \omega , CPC^\dagger$ up to a global phase $\omega \in {\pm1, \pm i}$.
\item additivity: $M_q(\ket{\Phi} \otimes \ket{\psi}) = M_q(\ket{\Phi}) + M_q(\ket{\psi})$.
\end{itemize}
Clifford invariance ensures that the SRE remains unchanged under the action of Clifford gates. In the resource theory of non-stabilizerness, the convex hull $\textsf{STAB} \equiv \{ \sum_i p_i \ket{\sigma_i} \bra{\sigma_i}: p_i > 0,, \sum_i p_i = 1 \}$ of pure stabilizer states $\ket{\sigma_i}$ defines the set of free states.
The free operations—referred to as stabilizer protocols—are completely positive trace-preserving (CPTP) maps that leave the set $\textsf{STAB}$ invariant. These protocols consist of the following elementary operations: i) action of Clifford unitaries; (ii) partial trace; (iii) measurements in the computational basis; (iv) composition with auxiliary qubits in state $\ket{0}$; (v) operations (i)--(iv) conditioned on the measurement outcomes and classical randomness.

The SRE for every integer $q \geq 2$ has been shown to be a pure-state stabilizer monotone~\cite{Bittel25operational}. That is, for any stabilizer protocol $\mathcal{E}$ such that $\mathcal{E}(\ket{\psi}\bra{\psi}) = \ket{\phi}\bra{\phi}$, the SRE is non-increasing:
\begin{equation}
M_q( \ket{\psi} ) \geq M_q(\ket{\phi}),
\label{eq:mono}
\end{equation}
provided that $q \geq 2$ is an integer. In contrast, for $0 < q < 2$, $M_q$ may increase under stabilizer protocols~\cite{Haug23monotones}.
The SRE admits an operational interpretation in the context of quantum property testing~\cite{Bittel25operational} and is bounded from above by other non-stabilizerness measures such as the robustness of magic~\cite{leone2022stabilizerrenyientropy} and the stabilizer extent~\cite{Bravyi16improved}. These alternative measures, however, involve minimization procedures that become computationally intractable beyond just a few qubits.

\subsection{Mana for qutrits}
For qutrits ($d=3$)~\footnote{The construction below applies to any odd prime dimension $d$.}, we define shift and clock operators as 
\begin{equation} \label{eq:clock_and_shift}
X = \sum_{k=0}^{d-1} |k+1\rangle \langle k | \quad \textrm{and} \quad Z=\sum_{k=0}^{d-1} \omega^k_d |k\rangle \langle k |,
\end{equation}
where $\omega_d=e^{i2\pi /d}$ and the addition is defined modulo $d$.
The generalized Pauli operators, also known as the Heisenberg-Weyl operators, are defined as
\begin{equation} \label{eq:pauli_qudit}
    T_{ab} = \omega_d^{-2^{-1}ab}Z^a X^{b}
\end{equation}
for $a,b \in \mathbb{Z}_d$ and with $2^{-1}$ the inverse element of $2$ in $\mathbb{Z}_d$. For a system of $N$ qutrits, the generalized Pauli strings read
\begin{equation}
    T_{\mathbf{u}}=T_{u_1,u_1'} \otimes  T_{u_2,u_2'} \otimes ... \otimes T_{u_N, u_N'},
\end{equation}
where $\mathbf{u} = (u_1, u_1', \ldots, u_N, u_N') \in \mathbb{Z}^{2N}_d$.
The Clifford group $\mathcal{C}_{N,d}$ maps, up to a phase, generalized Pauli strings to generalized Pauli strings, and stabilizer states of qutrits are obtained by action of $\mathcal{C}_{N,d}$ on $\ket{0}^N$.

The phase-space point operators, defined in terms of the Pauli strings as 
\begin{equation} \label{eq:phase-space}
    A_{\mathbf{0}} = \frac{1}{d^N} \sum_{\mathbf{u} \in \mathbb{Z}^{2N}_d } T_\mathbf{u},  \quad  A_{\mathbf{u}} = T_{\mathbf{u}}  A_{\mathbf{0}} T_{\mathbf{u}}^\dagger,
\end{equation}
are orthogonal, i.e, $\mathrm{tr}(A_\mathbf{u} A_\mathbf{u'}) = d^N \delta(\mathbf{u},\mathbf{u'})$, and thus they provide an orthogonal basis for the density matrix $\rho$ of a quantum state as 
\begin{equation}
    \rho = \sum_{\mathbf{u}\in \mathbb{Z}^{2N}_d }  W_{\rho} (\mathbf{u}) A_{\mathbf{u}}.
\end{equation}
where $W_{\rho} (\mathbf{u}) = \frac{1}{d^N} \mathrm{tr}(A_{\mathbf{u}}\rho)$ is the discrete Wigner function~\cite{gross2006hudsonstheoremfor} with the normalization condition $\sum_{\mathbf{u}}W_{\rho} (\mathbf{u}) =1$.
Finally, mana~\cite{Veitch2014theresourcetheory}, defined for qutrits in terms of the Wigner function as
\begin{equation} \label{eq:mana}
    \mathcal{M}(\rho) = \log_2 \left[ \sum_{\mathbf{u} \in \mathbb{Z}^{2N}_3}| W_{\rho}(\mathbf{u})| \right],
\end{equation}
 measures the negativity of the Wigner representation of $\rho$.

Mana is a faithful measure of non-stabilizerness for qutrit systems. Indeed, by the discrete Hudson theorem, a pure state \(\rho=\ket{\psi}\!\bra{\psi}\) has a nonnegative discrete Wigner function, \(W_{\ket{\psi}\bra{\psi}}(\mathbf{u})\ge 0\), if and only if it is a stabilizer state~\cite{gross2006hudsonstheoremfor}. Mana further satisfies the basic requirements of Clifford invariance and additivity, and it is a monotone not only for pure states [cf.\eqref{eq:mono}] but also for mixed states~\cite{Veitch2014theresourcetheory}. Notably, mana is the only known non-stabilizerness monotone whose definition avoids any minimization procedure for both pure and mixed states.

In the following, we will be considering calculation of mana for pure states, \( \rho = \ket{\psi}\bra{\psi} \), for which Eq.~\eqref{eq:mana} reduces to
\begin{equation} \label{eq:mana2}
    \mathcal{M}(\ket{\psi}) = \log_2 \left[  \, \sum_{A \in A_N}\frac{| \bra{\psi} A \ket{\psi} |  }{3^N}   \right] = \log_2 \left[  \, \sum_{ \mathbf{u} \in \mathbb{Z}^{2N}_3 }\frac{| \bra{\psi} A_{\mathbf{u}} \ket{\psi} |  }{3^N}   \right] ,
\end{equation}
where we have introduced the set $A_N =\{ A_{\mathbf{u}} \}_{\mathbf{u}\in \mathbb{Z}^{2N}_3}$ of $3^{2N}$ phase-space point operators~\eqref{eq:phase-space}, to highlight the similarity between the definition of mana for pure states and the SRE~\eqref{eq:sre}.

\section{Algorithms for computation of SRE and mana}
\label{sec:exact}
Equations~\eqref{eq:sre} and~\eqref{eq:mana2} show that evaluating the SRE and the mana requires computing \(\mathcal{N}^2=d^{2N}\) expectation values of Pauli strings \(P\) and phase-space point operators \(A\), respectively. While conceptually straightforward, this quickly becomes computationally prohibitive with the increase of the system size $N$. In this section, we present algorithms for the efficient evaluation of the SRE and mana for quantum states represented as state vectors.

\subsection{Calculation of SRE for qubits}

\subsubsection{Naive SRE calculation}

A direct numerical evaluation of the SRE in Eq.~\eqref{eq:sre} requires computing
\begin{equation}
\label{eq:sq}
S_q =
 \sum_{ \mathbf a, \mathbf b \in \mathbb{Z}_2^N} \langle \psi | P_{\mathbf{a},\mathbf{b}} | \psi \rangle^{2q},
\end{equation}
i.e., a sum over \(4^N\) expectation values of Pauli strings \(P_{\mathbf{a},\mathbf{b}}\) in the state \(\ket{\psi}\).
A straightforward approach is to generate the bit strings \(\mathbf{a}=(a_j)\) and \(\mathbf{b}=(b_j)\) in lexicographic order and, for each pair \((\mathbf{a},\mathbf{b})\), compute \(\ket{\phi}= P_{\mathbf{a},\mathbf{b}} \ket{\psi}\) and then the overlap \(\langle\psi|\phi\rangle\).
In practice, forming \( P_{\mathbf{a},\mathbf{b}} \ket{\psi}\) may require applying up to \(2N\) single-qubit operators \(X_j\) and \(Z_j\), represented as sparse matrices.
Generating \(\mathbf{a}\) and \(\mathbf{b}\) in Gray-code order~\cite{knuth1997art}, so that successive bit strings differ only by a single bit, reduces this cost: each step requires applying only a single \(X_j\) or \(Z_j\) operator.
This leads to the following algorithm.

\begin{Algorithm}{Naive calculation of SRE, $M_q(\ket{\psi})$, via Gray-code sweep of Pauli strings.\label{alg:sregray}\vspace{-0.2cm}}
\Require number of qubits \(N\); R\'enyi index \(q\neq 1\), state $|\psi\rangle$
\State \(S_q \gets 0\), \(\mathbf{a} \gets (0,\dots,0)\), \(\mathbf{b} \gets (0,\dots,0)\),
\State $\ket{\phi} \gets \ket{\psi}$ \Comment{initial Pauli string is the identity; at each step $\ket{\phi}= P \ket{\psi}$}
\For{\(k = 0\) to \(2^N-1\)}
  \For{\(l = 0\) to \(2^N-1\)}
    \State \(S_q \gets S_q + \langle \psi | \phi \rangle^{2q}\)
    \State \(\mathbf{b},t \gets \textsc{NextGray}_2(\mathbf{b})\) \Comment{Advance Gray code for \(\mathbf b\); \(t\) is the flipped bit.}
    \State $\ket{\phi} \gets Z_t \ket{\phi}$
  \EndFor
  \State \(\mathbf{a},w \gets \textsc{NextGray}_2(\mathbf{a})\) \Comment{Advance Gray code for \(\mathbf a\); \(w\) is the flipped bit.}
  \State $\ket{\phi} \gets X_w \ket{\phi}$
\EndFor
\State \Return \( M_q(\ket{\psi}) = \frac{1}{1-q} \log_2( S_q / 2^N)\).
\vspace{-0.2cm}
\end{Algorithm}

Since two Pauli strings \(P\) and \(Q\) either commute (\(PQ=QP\)) or anticommute (\(PQ=-QP\)), the state \(\ket{\phi}\) produced at each step of the algorithm is, up to an overall sign \(\pm 1\), equal to
$\ket{\phi}= P_{\mathbf{a},\mathbf{b}}\ket{\psi}$.
The above procedure requires evaluating \(4^N\) overlaps \(\langle \psi | \phi \rangle\), each at computational cost \(O(2^N)\), as well as performing \(4^N\) single-Pauli updates of the state,
\(
\ket{\phi}\leftarrow Z_t\ket{\phi}\) or 
\( \ket{\phi}\leftarrow X_w\ket{\phi},
\)
each also costing \(O(2^N)\).
Consequently, the overall time complexity of Algorithm~\ref{alg:sregray} is \(O(8^N)\), while the memory footprint is \(O(2^N)\), set by storing the state vectors \(\ket{\psi}\) and \(\ket{\phi}\).

\vspace{1.5cm}

\subsubsection{Efficient algorithm for exact SRE calculation}
\label{subsec:effSRE}
Each Pauli string \(P_{\mathbf{a},\mathbf{b}}\) appearing in the SRE
calculation can be written as a product \(X_{\mathbf a} Z_{\mathbf b}\),
where \(X_{\mathbf a} = \prod_{j=1}^N X^{a_j}_j\) contains only the
\(X_j\) operators and \(Z_{\mathbf b} = \prod_{j=1}^N Z^{b_j}_j\) contains
only the \(Z_j\) operators.
A key ingredient behind the exponential speedup over
Algorithm~\ref{alg:sregray} is that, for a fixed \(X\)-string
\(X_{\mathbf a}\), the entire inner loop over all \(Z\)-strings
\(Z_{\mathbf b}\) can be replaced by a single fast Hadamard transform.
In Algorithm~\ref{alg:sregray}, the inner Gray-code sweep updates
\(\mathbf b\in \mathbb{Z}_2^N\), applies \(\ket{\phi}\leftarrow Z_t\ket{\phi}\),
and accumulates the overlaps
\(
  \langle\psi|\phi\rangle^{2q}
  = \langle\psi | X_{\mathbf{a}} Z_{\mathbf{b}} | \psi\rangle^{2q}.
\)
Equivalently, denoting \(\ket{\psi'}= X_{\mathbf a}\ket{\psi}\), the
inner loop computes
\(\chi_{\mathbf b}\equiv \bra{\psi'}Z_{\mathbf b}\ket{\psi}\) for all
\(\mathbf b\in \mathbb{Z}_2^N\).

Expanding \(\ket{\psi'}=\sum_{\mathbf x}\beta_{\mathbf x}\ket{\mathbf x}\)
and \(\ket{\psi}=\sum_{\mathbf x}\alpha_{\mathbf x}\ket{\mathbf x}\) in
the computational basis \(\{\ket{\mathbf x}\}\) with
\(\mathbf x\in\mathbb{Z}_2^N\), and using
\(Z_{\mathbf b}\ket{\mathbf x}=(-1)^{\mathbf b\cdot\mathbf x}\ket{\mathbf x}\),
we obtain the explicit expression
\begin{equation}
\label{eq:Z_FT_qubits_section}
\chi_{\mathbf b}
=
\sum_{\mathbf x\in \mathbb{Z}_2^N }
\beta_{\mathbf x}^*\alpha_{\mathbf x}\,(-1)^{\mathbf b\cdot\mathbf x}.
\end{equation}
Thus, the length-\(2^N\) vector \((\chi_{\mathbf b})_{\mathbf b \in \mathbb{Z}_2^N}\)
is the Walsh--Hadamard transform (i.e., the Fourier transform over
\(\mathbb{Z}_2^N\)) of the vector with components
\(v_{\mathbf x}\equiv \beta_{\mathbf x}^*\alpha_{\mathbf x}\).
In matrix form,
\begin{equation}
\label{eq:Z_ovN_simplified}
(\chi_{\mathbf b})_{\mathbf b \in \mathbb{Z}_2^N}
=
H_2^{\otimes N}\,(v_{\mathbf x})_{\mathbf x \in \mathbb{Z}_2^N },
\end{equation}
where \(H_2^{\otimes N}\) is the \(N\)-fold tensor product of the
unnormalized single-qubit Hadamard matrix
\(
  H_2
  =
  \begin{pmatrix}
    1 & 1 \\
    1 & -1
  \end{pmatrix}.
\)
Computing \((\chi_{\mathbf b})_{\mathbf b  \in \mathbb{Z}_2^N}\) by explicit
matrix multiplication would require a computational cost \(O(4^N)\), whereas
the fast Hadamard transform evaluates Eq.~\eqref{eq:Z_ovN_simplified} in
\(O(N2^N)\) time~\cite{Fino76}.
Since this computation must be repeated for all \(2^N\) choices of
\(\mathbf a\), the resulting overall cost is \(O(N4^N)\), yielding an
exponential improvement over the \(O(8^N)\) scaling of
Algorithm~\ref{alg:sregray}.
This leads to the following algorithm for the numerically exact calculation
of the SRE.

\begin{Algorithm}{Calculation of the SRE \(M_q(\ket{\psi})\) using the fast Hadamard transform.
\label{alg:sre_eff}\vspace{-0.2cm}}
\Require Number of qubits \(N\); R\'enyi index \(q\neq 1\); state \(\ket{\psi}\).
\State \(S_q \gets 0\), \(\mathbf a \gets (0,\dots,0) \in \mathbb{Z}_2^N \), \(\ket{\psi'} \gets \ket{\psi}\) \Comment{\(\ket{\psi'}=\bigl(\prod_{j=1}^N X_j^{a_j}\bigr)\ket{\psi}\).}
\For{\(k = 0\) to \(2^N-1\)}
  \State Extract amplitudes \(\beta_{\mathbf x}\) of \(\ket{\psi'}=\sum_{\mathbf x}\beta_{\mathbf x}\ket{\mathbf x}\) and \(\alpha_{\mathbf x}\) of \(\ket{\psi}=\sum_{\mathbf x}\alpha_{\mathbf x}\ket{\mathbf x}\).
  \State Form \(v_{\mathbf x}\gets \beta_{\mathbf x}^*\alpha_{\mathbf x}\) for all \(\mathbf x\in \mathbb{Z}_2^N\).
  \State \((\chi_{\mathbf b})_{\mathbf b} \gets \textsc{FastHadamardTransform}\bigl((v_{\mathbf x})_{\mathbf x}\bigr)\).
  \State \(S_q \gets S_q + \sum_{\mathbf b\in \mathbb{Z}_2^N} \chi_{\mathbf b}^{2q}\)
  \Comment{Accumulates \(\sum_{\mathbf b}\langle\psi'|Z_{\mathbf b}|\psi\rangle^{2q}\).}
  \State \(\mathbf a, t \gets \textsc{NextGray}_2(\mathbf a)\) \Comment{Advance Gray code for \(\mathbf a\); \(t\) is the flipped bit.}
  \State \(\ket{\psi'} \gets X_t \ket{\psi'}\).
\EndFor
\State \Return \( M_q(\ket{\psi})=\frac{1}{1-q}\log_2 (S_q/2^N) \).
\vspace{-0.2cm}
\end{Algorithm}

Algorithm~\ref{alg:sre_eff} has time complexity \(O(N4^N)\) and memory footprint \(O(2^N)\), set by storing the state vectors \(\ket{\psi}\), \(\ket{\psi'}\), and a work array of length \(2^N\). Moreover the loop over $k$ can be split into chunks, each evaluated separately, which enables a high degree of parallelism, see Sec.~\ref{subsec:exactSRE}.

\subsection{Monte Carlo sampling algorithm for SRE}
\label{subsec:SREmc}

\subsubsection{Direct sampling of Pauli strings}

Treating the Pauli strings \( P \in \mathcal{P}_N\) as a basis of the operator
space, any pure state \(\ket{\psi}\) of \(N\) qubits with density matrix
\(\rho = \ket{\psi}\!\bra{\psi}\) admits the expansion
\(\rho = 2^{-N} \sum_{P\in\mathcal{P}_N} \braket{\psi|P|\psi}\,P\).
The normalization of the state, \(\|\ket{\psi}\| = 1\), then implies that
\(\sum_{P\in\mathcal{P}_N} \braket{\psi|P|\psi}^2 / 2^N = 1\), which leads to
a natural probability distribution over Pauli strings,
\begin{equation}
\label{eq:distNAT}
  \pi(P) \;=\; \frac{\braket{\psi|P|\psi}^2}{2^N},
  \qquad P\in\mathcal{P}_N.
\end{equation}
Sampling Pauli strings according to the distribution \(\pi(P)\) enables a
Monte Carlo estimation of the SRE.
Given a sample \(\{P_k\}_{k=1}^{n_S}\) drawn from the distribution \(\pi\),
a simple estimator of \(S_q\) in Eq.~\eqref{eq:sq} is
\begin{equation}
  \hat{S}_q(\ket{\psi})
  \;=\;
  \frac{1}{n_S}\sum_{k=1}^{n_S} \braket{\psi|P_k|\psi}^{2(q-1)},
\end{equation}
see, e.g.,~\cite{lami2023perfect},
so that a sampling-based estimate of the SRE reads
\(
  \hat{M}_q(\ket{\psi})
  =
  -\log_2 [\hat{S}_q(\ket{\psi})].
\)
The distribution \(\pi(P)\) can be sampled, for example, directly with a
Metropolis--Hastings algorithm~\cite{Turkeshi25spectrum}, via Bell
sampling~\cite{Tarabunga25}, or using sampling approaches adapted to tensor
network states~\cite{tarabunga23gauge,lami2023perfect}.
For large \(n_S\), the central limit theorem gives the statistical uncertainty
\(\delta S_q \sim \sigma_b / \sqrt{n_S}\), where
\(\sigma_b^2 = \mathrm{Var}_\pi\big[\braket{\psi|P|\psi}^2\big]\).
Fixing \(q>1\), we see that the uncertainty on \(M_q\) is controlled by
\begin{equation}
\label{eq:M2mNcerror}
  \delta M_q
  \;\approx\;
  \frac{1}{\ln 2}\,\frac{\delta S_q}{S_q}
  \;\sim\;
  \frac{\sigma_b}{S_q \sqrt{n_S}}.
\end{equation}
For non-stabilizer states, \(S_q(\ket{\psi})\) can decay
exponentially with the number of qubits, so that the statistical error
\(\delta M_q\) for fixed \(n_S\) can grow, in the worst-case scenario,
exponentially with system size \(N\).
In such cases, achieving a fixed additive precision in \(M_q\) for any
\(q>1\) requires a number of samples \(n_S\) that is exponential in \(N\),
rendering the naive sampling approach inefficient~\footnote{For \(q=1\),
the variance is instead at most polynomial in \(N\), enabling an efficient
estimation of \(M_1(\ket{\psi})\).}.

\subsubsection{Fast-Hadamard-enhanced sampling.}
The sampling of Eq.~\eqref{eq:distNAT} treats all Pauli strings on equal
footing.
Here we exploit the fast Hadamard transform to accelerate the sampling
procedure.
The key observation is that, for a Pauli string parametrized as
\(
P = X_{\mathbf a} Z_{\mathbf b},
\)
a single fast Hadamard transform over the \(Z\)-strings yields the
expectation values
\(\braket{\psi|X_{\mathbf a} Z_{\mathbf b}|\psi}\) for all Pauli strings
\(X_{\mathbf a} Z_{\mathbf b}\) at once, for any fixed \(X\)-string
\(X_{\mathbf a}\), an idea already used in Algorithm~\ref{alg:sre_eff}.

Following similar approaches for mana~\cite{tarabunga2023critical} and
entanglement~\cite{Boer19,Hong10,Schliemann11,Zaletel11},
for each fixed \(X_{\mathbf a}\) we define a regularized ``energy''
\begin{equation}
\label{eq:feP}
  f_\varepsilon(X_{\mathbf a})
  \;=\;
  -\log_2\!\Biggl(
    \sum_{\mathbf b \in \mathbb{Z}_2^N}
    \braket{\psi|X_{\mathbf a} Z_{\mathbf b}|\psi}^4
    \;+\;\varepsilon
  \Biggr),
  \qquad \varepsilon>0,
\end{equation}
so that the inner sum is strictly positive for all \(\mathbf a  \in \mathbb{Z}_2^N\).
The parameter \(\varepsilon \ll 1\) controls a small additive bias,
ensuring that \(f_\varepsilon(X_{\mathbf a})\) remains finite even if
\(\sum_{\mathbf b \in \mathbb{Z}_2^N}
    \braket{\psi|X_{\mathbf a} Z_{\mathbf b}|\psi}^4\) vanishes for certain
\(\mathbf a\).
For a generic non-stabilizer state, \(\varepsilon=0\) is sufficient.
To sample \(X_{\mathbf a}\) we introduce the Boltzmann distribution
\begin{equation}
\label{eq:Boltz}
  \Pi_\beta(X_{\mathbf a})
  \;=\;
  \frac{1}{Z_\beta^{(\varepsilon)}}\,
  e^{-\beta f_\varepsilon(X_{\mathbf a})},
  \qquad
  Z_\beta^{(\varepsilon)}
  =
  \sum_{\mathbf a} e^{-\beta f_\varepsilon(X_{\mathbf a})},
\end{equation}
with an auxiliary inverse temperature \(\beta\in[0,1]\).
The expectation value of \(f_\varepsilon\) with respect to \(\Pi_\beta\) is
\begin{equation}
  \label{eq:f_expect_beta_reg}
  \bigl\langle f_\varepsilon(X_{\mathbf a})\bigr\rangle_\beta
  =
  \sum_{\mathbf a  \in \mathbb{Z}_2^N } \Pi_\beta(X_{\mathbf a})\,f_\varepsilon(X_{\mathbf a})
  =
  -\frac{d}{d\beta}\log Z_\beta^{(\varepsilon)},
\end{equation}
and integrating over \(\beta\) yields the standard thermodynamic
identity
\begin{equation}
  \label{eq:TI_logZ_reg}
  \int_0^1 d\beta\,
  \bigl\langle f_\varepsilon(X_{\mathbf a})\bigr\rangle_\beta
  =
  \log Z_1^{(\varepsilon)} - \log Z_0^{(\varepsilon)}.
\end{equation}
At \(\beta=1\) we obtain
\begin{equation}
  Z_1^{(\varepsilon)}
  =
  \sum_{\mathbf a} e^{-f_\varepsilon(X_{\mathbf a})}
  =
  \sum_{\mathbf a, \mathbf b \in  \mathbb{Z}_2^N}
  \braket{\psi|X_{\mathbf a} Z_{\mathbf b}|\psi}^4
  + 2^N \varepsilon
  =
  S_2(\ket{\psi}) + 2^N \varepsilon,
\end{equation}
where we used that each Pauli string can be written uniquely as
\(X_{\mathbf a} Z_{\mathbf b}\).
At \(\beta=0\), all Boltzmann weights are equal to unity and
\(
  Z_0^{(\varepsilon)}  = 2^N.
\)
Combining these relations with Eq.~\eqref{eq:TI_logZ_reg}
and denoting
\begin{equation}
\label{eq:Ie}
  I_\varepsilon(\psi)
  \;=\;
  \int_0^1 d\beta\,
  \bigl\langle f_\varepsilon(X_{\mathbf a})\bigr\rangle_\beta
\end{equation}
yields the SRE as 
\begin{equation}
  \label{eq:M2_TI_reg_explicit_correct}
  M_2(\ket{\psi})
  =
  -\log_2\!\bigl(e^{I_\varepsilon(\psi)} - \varepsilon\bigr).
\end{equation}

\subsubsection{Monte Carlo estimator and sampling error.}

In our numerical approach, to calculate the SRE, we discretize the integral~\eqref{eq:Ie} over \(\beta\) on a grid \(\{\beta_\ell\}_{\ell=1}^{L}\) with quadrature weights \(\{w_\ell\}\),
\begin{equation}
  I_\varepsilon(\psi)
  \;\approx\;
  \hat I_\varepsilon(\psi)
  =
  \sum_{\ell=1}^L w_\ell\,
  \widehat{\bigl\langle f_\varepsilon\bigr\rangle}_{\beta_\ell},
\end{equation}
and estimate \(\langle f_\varepsilon\rangle_{\beta_\ell}\) by running a
Markov chain at fixed \(\beta_\ell\) with stationary distribution
\(\Pi_{\beta_\ell}(X_{\mathbf a})\).
If the chain at \(\beta_\ell\) produces a series of samples
\(\{X_{\mathbf a}^{(n,\ell)}\}_{n=1}^{n_\ell}\), we
use the sample mean
\begin{equation}
  \widehat{\bigl\langle f_\varepsilon\bigr\rangle}_{\beta_\ell}
  =
  \frac{1}{n_\ell}\sum_{t=1}^{n_\ell}
  f_\varepsilon\bigl(X_{\mathbf a}^{(n,\ell)}\bigr).
\end{equation}

For each \(\beta_\ell\), the central limit theorem for Markov chains implies
that the variance of the sample mean
\(\widehat{\langle f_\varepsilon\rangle}_{\beta_\ell}\) scales as
\( 2\,\tau_{\mathrm{int}}(\beta_\ell)\sigma_f^2(\beta_\ell)/n_\ell\),
where \(\sigma_f^2(\beta_\ell)\) is the variance of
\(f_\varepsilon(X_{\mathbf a})\) under \(\Pi_{\beta_\ell}\) and
\(\tau_{\mathrm{int}}(\beta_\ell)\) is the corresponding integrated
autocorrelation time. Since the runs for different
\(\beta_\ell\) are statistically independent, the variance of \(\hat I_\varepsilon\)
is 
\begin{equation}
  \label{eq:Var_I_eps}
  \mathrm{Var}\bigl[\hat I_\varepsilon(\psi)\bigr]
  \;\approx\;
  \sum_{\ell=1}^L
  w_\ell^2\,
  \frac{2\,\tau_{\mathrm{int}}(\beta_\ell)}{n_\ell}\,
  \sigma_f^2(\beta_\ell).
\end{equation}

The role of the regularization parameter \(\varepsilon\) is to ensure that
the logarithm in \(f_\varepsilon(X_{\mathbf a})\) is well-defined even when
\(
S(\mathbf a)
=
\sum_{\mathbf b}\braket{\psi|X_{\mathbf a}Z_{\mathbf b}|\psi}^4
\)
vanishes for some \(\mathbf a\).
If we know a priori that \(S(\mathbf a)>0\) for all \(\mathbf a\)---as is
empirically the case for the quantum-circuit states analyzed in
Sec.~\ref{subsec:uses}---we may safely take the limit
\(\varepsilon\to 0\) in Eqs.~\eqref{eq:feP}
and~\eqref{eq:M2_TI_reg_explicit_correct} and work with
\(f(X_{\mathbf a}) = -\log S(\mathbf a)\).
In this regime \(f(X_{\mathbf a})\) is finite for all \(\mathbf a\).

A notable exception arises for states with U(1) symmetry (conserved total magnetization \(\sum_j Z_j\)) or with a \(\mathbb{Z}_2\) symmetry generated by \(\prod_j Z_j\) (conserved computational-basis parity): for both types, \(\braket{\psi|X_{\mathbf a}Z_{\mathbf b}|\psi}=0\) for every \(\mathbf b\) whenever \(|\mathbf a|\) is odd, since the basis states of \(|\psi\rangle\) all share the same Hamming-weight parity and \(X_{\mathbf a}\) with an odd number of flips maps them to the opposite parity, rendering \(X_{\mathbf a}|\psi\rangle\) orthogonal to \(|\psi\rangle\).
All \(2^{N-1}\) odd-weight \(X\)-sectors therefore contribute nothing to the SRE and \(S(\mathbf a)=0\) for those sectors.
For a \(\mathbb{Z}_2\) symmetry generated by \(\prod_j X_j\), the constraint instead falls on \(\mathbf b\)-strings: \(\braket{\psi|X_{\mathbf a}Z_{\mathbf b}|\psi}=0\) for \(\mathbf b\) of the wrong parity, but no entire \(X\)-sector is zeroed out, so \(\varepsilon=0\) and an unrestricted chain remain valid.
More generally, symmetries must be analyzed case by case. When the structure is known, the preferred approach is to restrict the Markov chain to the non-vanishing \(X\)-sectors (e.g., even-weight \(\mathbf a\) for U(1) or \(\mathbb{Z}_2^Z\) symmetry); setting \(\varepsilon>0\) is the appropriate fallback when the symmetry is unknown or not explicitly exploited, though it comes at the cost of the chain occasionally visiting zero-sum sectors (with Boltzmann weight \(\propto\varepsilon^\beta\)), which reduces sampling efficiency relative to the symmetry-aware approach.

Denoting \(s_{\min} = \min_{\mathbf a} S(\mathbf a)\), the variance
\(\sigma_f^2(\beta_\ell)\) of \(f(X_{\mathbf a})\) under
\(\Pi_{\beta_\ell}\) is then bounded from above by a constant of order
\(\log^2(1/s_{\min})\).

For the quantum-circuit states considered later in this work
(Sec.~\ref{subsec:uses}), we empirically find
\(
s_{\min}(N) \sim e^{-cN}
\)
with some constant \(c>0\).
In this case \(\log(1/s_{\min}) \sim cN\), so that the worst-case variance
of \(f(X_{\mathbf a})\) scales at most quadratically with system size,
\(\sigma_f^2(\beta_\ell) = O(N^2)\), and the corresponding contribution to
\(\mathrm{Var}[\hat I(\psi)]\) in Eq.~\eqref{eq:Var_I_eps} grows at most
polynomially in \(N\).

For \(\varepsilon=0\), the stabilizer Rényi entropy in
Eq.~\eqref{eq:M2_TI_reg_explicit_correct} can be estimated as
\(\hat M_2(\ket{\psi}) = -\hat I(\psi)/\ln 2\).
Using the variance estimate
\begin{equation}
  \label{eq:Var_I_noeps_samples}
  \mathrm{Var}\bigl[\hat I(\psi)\bigr]
  \;\approx\;
  \sum_{\ell=1}^L
  w_\ell^2\,
  \frac{2\,\tau_{\mathrm{int}}(\beta_\ell)}{n_\ell}\,
  \sigma_f^2(\beta_\ell),
\end{equation}
where \(n_\ell\) is the number of Monte Carlo samples at \(\beta_\ell\),
\(\tau_{\mathrm{int}}(\beta_\ell)\) the integrated autocorrelation time,
and \(\sigma_f^2(\beta_\ell)\) the variance of \(f(X_{\mathbf a})\) under
\(\Pi_{\beta_\ell}\), error propagation yields the statistical uncertainty
on the SRE,
\begin{equation}
\label{eq:m2error}
  \delta M_2
  \equiv
  \sqrt{\mathrm{Var}\bigl[\hat M_2(\ket{\psi})\bigr]}
  \;\approx\;
  \frac{1}{\ln 2}\,
  \sqrt{
    \sum_{\ell=1}^L
    w_\ell^2\,
    \frac{2\,\tau_{\mathrm{int}}(\beta_\ell)}{n_\ell}\,
    \sigma_f^2(\beta_\ell)
  }.
\end{equation}
Equivalently, introducing an effective number of independent samples
\(n_{\mathrm{eff}} \sim \sum_\ell n_\ell / \tau_{\mathrm{int}}(\beta_\ell)\),
we have the scaling
\(\delta M_2 \propto 1/\sqrt{n_{\mathrm{eff}}}\), up to polynomial factors
in \(N\) coming from \(\sigma_f^2(\beta_\ell)\).

A polynomial growth of \(\delta M_2\) with \(N\) is in stark contrast to
the exponential increase of the statistical error in the naive sampling
scheme, cf.~Eq.~\eqref{eq:M2mNcerror}.
Consequently, for generic quantum states with \(S(\mathbf a)>0\), the
present thermodynamic-integration approach requires only a polynomial
increase of the total sampling effort (i.e.\ the number of Markov-chain
updates \(\{n_\ell\}\) at each \(\beta_\ell\)) in order to maintain a fixed
statistical precision \(\delta M_2\) on the SRE as the
system size grows.

In the general case, \(S(\mathbf a)\) may vanish for some \(\mathbf a\).
Then, the limit $\varepsilon \to 0$ is singular, and the error propagation yields
\begin{equation}
  \label{eq:deltaM2_eps_general}
  \delta M_2
  \equiv
  \sqrt{\mathrm{Var}\bigl[\hat M_2(\ket{\psi})\bigr]}
  \;\approx\;
  \frac{1}{\ln 2}\,
  \Biggl(1 + \frac{\varepsilon}{S_2(\ket{\psi})/2^N}\Biggr)\,
  \sqrt{\mathrm{Var}\bigl[\hat I_\varepsilon(\psi)\bigr]}.
\end{equation}
If \(S_2(\ket{\psi})/2^N \lesssim \varepsilon\), the prefactor becomes large,
\(1 + \varepsilon/(S_2/2^N) = O\!\bigl(\varepsilon/(S_2/2^N)\bigr)\),
and the uncertainty \(\delta M_2\) is strongly amplified by the finite
value of \(\varepsilon\).
In this regime, resolving \(M_2\) with fixed relative precision
would require reducing \(\varepsilon\) in tandem with
\(S_2(\ket{\psi})/2^N\), which in turn increases the variance of
\(f_\varepsilon\) as \(\sim \log^2(1/\varepsilon)\), and in the worst-case
scenario may lead to an exponential increase of \(\delta M_2\).
However, the vanishing of \(S(\mathbf a)\) for a fixed set of
\(\{\mathbf a\}\) is a fine-tuned property of the quantum state
\(\ket{\psi}\), and does not occur in most of the practical applications
of the sampling scheme.

% \Statex
% \State \textbf{Assumption:} a routine \(\textsc{EvalEnergy}(P_{\mathbf a})\) is available, returning
% \[
%   f_\varepsilon(P_{\mathbf a})
%   = -\log\!\Bigl(
%       \sum_{\mathbf b}
%       \braket{\psi|P_{\mathbf a}P_{\mathbf b}|\psi}^4
%       + \varepsilon
%     \Bigr).
% \]
% \State This routine internally evaluates the \(\mathbf b\)-sum using a fast Hadamard transform.

\subsubsection{Sampling scheme for SRE.}
\label{subsec:sre_mc}

Here, for simplicity, we focus on the generic situation with \(S(\mathbf a)>0\) for all \(\mathbf a\) and fix \(\varepsilon = 0\).
To sample the Boltzmann probability distribution \(\Pi_\beta(X_{\mathbf a})\) in Eq.~\eqref{eq:Boltz}, we use the Metropolis--Hastings algorithm~\cite{Metropolis1953equation,Hastings70}, which builds a Markov chain
\(\{X_{\mathbf a}^{(n)}\}_{n\geq 0}\) with stationary distribution \(\Pi_\beta\).
Starting from an initial pattern \(X_{\mathbf a}^{(0)}\), each step consists of proposing a new pattern \(X_{\mathbf a'}\) from a symmetric proposal distribution (e.g.\ by flipping a randomly chosen subset of bits in \(\mathbf a\)), evaluating the corresponding energy difference
\(
\Delta f = f(X_{\mathbf a'}) - f(X_{\mathbf a}^{(n)}),
\)
and accepting the move with probability
\[
  p_{\mathrm{acc}}
  =
  \min\bigl\{1,\exp\bigl[-\beta\,\Delta f\bigr]\bigr\},
\]
otherwise keeping the current state.
After an initial thermalization period, successive samples along the chain provide an ergodic sampling of \(X_{\mathbf a}\) distributed according to \(\Pi_\beta\), allowing us to estimate \(\langle f\rangle_\beta\) by averaging over the sampled configurations.

The central computational step of our sampling scheme is the evaluation of
the ``energy'' \(f(X_{\mathbf a}) = -\log S(\mathbf a)\), which we make
efficient by exploiting the fast Hadamard transform.
For a given \(X\)-pattern \(X_{\mathbf a}\), the routine \textsc{EvalEnergy}
first computes the amplitudes of the states \(\ket{\psi}\) and
\(X_{\mathbf a}\ket{\psi}\) in the computational basis and forms a vector
\(v_{\mathbf x}\) whose components are suitable products of these amplitudes
(as in Algorithm~\ref{alg:sre_eff}).
A single fast Walsh--Hadamard transform applied to \(v\) then yields all
overlaps
\(\chi_{\mathbf b} = \braket{\psi|X_{\mathbf a}Z_{\mathbf b}|\psi}\) for
\(\mathbf b\in \mathbb{Z}_2^N\) in one shot, in time \(O(N2^N)\).
The quantity
\(S(\mathbf a) = \sum_{\mathbf b}\chi_{\mathbf b}^4\) is obtained by a
single pass over these overlaps, and \textsc{EvalEnergy} finally returns
\(f(X_{\mathbf a}) = -\log S(\mathbf a)\).
In this way, an explicit sum over all \(4^N\) Pauli strings is replaced by
a single fast transform plus linear-time post-processing.
This procedure is repeated for each \(\beta\) on a uniform grid in
\([0,1]\), generating \(n_S\) Monte Carlo samples at each \(\beta\).

\begin{Algorithm}{Sampling-based estimation of the SRE \(M_2(\ket{\psi})\).
\label{alg:sre_sampling_fht} \vspace{-0.2cm} }
\Require Number of qubits \(N\); state \(\ket{\psi}\);
         even integer \(L\) (number of grid points); 
         number of samples \(n_S\).

\State Set up a grid \(\beta_\ell = \ell\,\Delta\beta\), \(\ell=0,\dots,L-1\),
       with \(\Delta\beta = 1/(L-1)\), and Simpson weights \(w_\ell\).

\For{\(\ell = 0\) to \(L-1\)} \Comment{Independent Markov chains at each \(\beta_\ell\)}
  \State Initialize a random \(X\)-pattern \(X_{\mathbf a}^{(0,\ell)}\), and set \(A_\ell \gets 0\).
  \For{\(n = 1\) to \(n_S\)}
    \State Propose a new pattern \(X_{\mathbf a'}\) by a local move on \(\mathbf a\).
    \State Compute \(f(X_{\mathbf a'}) \gets \textsc{EvalEnergy}(X_{\mathbf a'})\), and let \(f(X_{\mathbf a}^{(n-1,\ell)})\) be the current value.
    \State Set
      \(p_{\mathrm{acc}} \gets
      \min\!\bigl\{1,\exp\bigl[-\beta_\ell\bigl(
          f(X_{\mathbf a'}) - f(X_{\mathbf a}^{(n-1,\ell)})
        \bigr)\bigr]\bigr\}\).
    \State With probability \(p_{\mathrm{acc}}\) set
           \(X_{\mathbf a}^{(n,\ell)} \gets X_{\mathbf a'}\); otherwise set
           \(X_{\mathbf a}^{(n,\ell)} \gets X_{\mathbf a}^{(n-1,\ell)}\).
    \State Accumulate
      \(A_\ell \gets A_\ell + f\bigl(X_{\mathbf a}^{(n,\ell)}\bigr)\).
  \EndFor
  \State Estimate \(\langle f\rangle_{\beta_\ell}\) as
    \(\widehat{\bigl\langle f\bigr\rangle}_{\beta_\ell}
    \gets A_\ell / n_S\).
\EndFor

\State \Return
  \(\hat M_2(\ket{\psi})
  = -\dfrac{1}{\ln 2}\,
  \sum_{\ell=0}^{L-1}
  w_\ell\,
  \widehat{\bigl\langle f\bigr\rangle}_{\beta_\ell}\).
  \vspace{-0.1cm} 
\end{Algorithm}

The overall complexity of the resulting Algorithm~\ref{alg:sre_sampling_fht} is dominated by calls to \textsc{EvalEnergy} inside the Markov chains.
Each evaluation costs \(O(N2^N)\) operations, and it is invoked once per Monte Carlo update at each \(\beta_\ell\).
Neglecting the thermalization phase, the dominating computational cost of the algorithm thus scales as $O\Bigl( L n_S \,N2^N\Bigr)$.
For fixed grid size \(L\) and polynomially growing sample numbers \(n_S\), the algorithm is therefore quasi-linear in the Hilbert-space dimension \(2^N\), in sharp contrast to any scheme that would explicitly iterate over all \(4^N\) Pauli strings.

\subsection{Calculation of mana for pure qutrit states}
\label{subsec:manaPure}

\subsubsection{Naive mana calculation}
The close formal resemblance between Eqs.~\eqref{eq:sre} and~\eqref{eq:mana2}
implies that the naive Gray-code sweep used for the SRE in
Algorithm~\ref{alg:sregray} can be adapted to compute mana with only minor
modifications.
For a pure state \(\ket{\psi}\) of \(N\) qutrits, the mana
\(\mathcal{M}(\ket{\psi})\) is obtained from the discrete Wigner function
and requires evaluating the \(9^N\) expectation values
\(\bra{\psi}A_{\mathbf u}\ket{\psi}\), where \(\{A_{\mathbf u}\}\) denotes
the set of phase-space point operators labeled by
\(\mathbf u\in\mathbb{Z}_3^{2N}\).

In the following, we set \(\omega\equiv \omega_3=e^{i2\pi/3}\) and
reparameterize \(\mathbf u\) as a pair of \(N\)-component ternary strings
\((\mathbf a,\mathbf b)\in\mathbb{Z}_3^N\times\mathbb{Z}_3^N\), defining
\begin{equation}
\label{eq:Aab_def}
A_{\mathbf a\mathbf b}
=
\Bigl(\prod_{j=1}^N X_j^{a_j} Z_j^{b_j}\Bigr)\,
A_{\mathbf 0}\,
\Bigl(\prod_{j=1}^N X_j^{a_j} Z_j^{b_j}\Bigr)^\dagger .
\end{equation}
Because \(X_j^{a}Z_j^{b}=\omega^{r(a,b)}Z_j^{b}X_j^{a}\) (with
\(r(a,b)\in\{0,1,2\}\)), this definition is equivalent to
Eq.~\eqref{eq:phase-space} up to an overall phase.
Moreover, a direct calculation shows that the origin operator acts as a
reflection in the computational basis,
\begin{equation}
\label{eq:A0_reflection}
A_{\mathbf 0}\ket{\mathbf x}=\ket{-\mathbf x},
\end{equation}
with \(\mathbf x\in\mathbb{Z}_3^N\) and subtraction taken modulo \(3\)~\cite{tarabunga2023critical}.

A straightforward (but costly) approach to compute mana is therefore to
iterate over all \((\mathbf a,\mathbf b)\in\mathbb{Z}_3^N\times\mathbb{Z}_3^N\),
construct \(A_{\mathbf a\mathbf b}\ket{\psi}\), and evaluate the overlaps
\(\bra{\psi}A_{\mathbf a\mathbf b}\ket{\psi}\).
Similarly to the qubit case, one can order the updates using a ternary Gray code
for \(\mathbf a\) and \(\mathbf b\) (so that successive strings differ at a
single position), resulting in Algorithm~\ref{alg:mana_gray}.
The memory cost of this scheme is \(O(3^N)\), set by storing state vectors
of \(N\) qutrits, while the runtime is dominated by the evaluation of the
\(9^N\) overlaps, each involving \(O(3^N)\) operations.
This leads to an overall computational complexity \(O(27^N)\), which grows
exponentially with system size and restricts this naive approach to
relatively small \(N\).

\begin{Algorithm}{Mana calculation via a Gray-code sweep of phase-space operators.
\label{alg:mana_gray}\vspace{-0.2cm}}
\Require Number of qutrits \(N\); state \(\ket{\psi}\).
\State \(m \gets 0\); \(\,\mathbf a\gets (0,\ldots,0)\), \(\mathbf b\gets (0,\ldots,0)\); \(\ket{\phi}\gets A_{\mathbf 0}\ket{\psi}\). \Comment{Initialize at \((\mathbf a,\mathbf b)=(\mathbf 0,\mathbf 0)\).}
\For{\(k=0\) to \(3^N-1\)} \Comment{Loop over \(\mathbf a\in\mathbb{Z}_3^N\).}
  \For{\(l=0\) to \(3^N-1\)} \Comment{Loop over \(\mathbf b\in\mathbb{Z}_3^N\).}
    \State \(m \gets m + \bigl|\bra{\psi}\phi\rangle\bigr|\).
    \State \(\mathbf b, t \gets \textsc{NextGray}_3(\mathbf b)\). \Comment{Ternary Gray step; \(t\) is the position incremented.}
    \State \(\ket{\phi} \gets Z_t\ket{\phi}\). \Comment{Update corresponding to \(\mathbf b\to \mathbf b'\).}
  \EndFor
  \State \(\mathbf a, w \gets \textsc{NextGray}_3(\mathbf a)\). \Comment{Advance ternary Gray code for \(\mathbf a\).}
  \State \(\ket{\phi} \gets X_w\ket{\phi}\). \Comment{Update corresponding to \(\mathbf a\to \mathbf a'\).}
\EndFor
\State \Return \( \mathcal{M}(\ket{\psi} = \log_2( m) \) 
\vspace{-0.2cm}
\end{Algorithm}

\subsubsection{Fast Hadamard transform for mana}
\label{subsec:manafast}
As in the qubit case, an exponential speedup over the naive Gray-code sweep
for mana hinges on accelerating the inner loop over the ``\(Z\)-type''
phase-space index.
Recall that mana for \(N\) qutrits requires evaluating the overlaps
\(\bra{\psi}A_{\mathbf a\mathbf b}\ket{\psi}\) for all
\((\mathbf a,\mathbf b)\in\mathbb{Z}_3^N\times\mathbb{Z}_3^N\).
For a fixed \(\mathbf a\), we define the \(\mathbf a\)-dependent state
\begin{equation}
\label{eq:def_psiprime_mana}
\ket{\psi_{\mathbf a}}
\equiv
\Bigl(\prod_{j=1}^N X_j^{a_j}\Bigr)\ket{\psi}.
\end{equation}
Then, up to an overall phase that is irrelevant for the absolute value
entering mana, the inner sweep over \(\mathbf b\) amounts to evaluating the
family of overlaps
\begin{equation}
\label{eq:mana_overlap_family}
\bra{\psi}\,A_{\mathbf a\mathbf b}\,\ket{\psi}
=
\bra{\psi_{\mathbf a}}\,Z_{\mathbf b}\,A_{\mathbf 0}\,Z_{\mathbf b}^\dagger\,\ket{\psi_{\mathbf a}},
\qquad
Z_{\mathbf b}\equiv \prod_{j=1}^N Z_j^{b_j},
\end{equation}
for all \(\mathbf b\in\mathbb{Z}_3^N\).
Using \(A_{\mathbf 0}\ket{\mathbf x}=\ket{-\mathbf x}\) (with arithmetic
modulo \(3\))~\cite{tarabunga2023critical}, these overlaps can be reduced to
a ternary Hadamard-type transform, i.e., a Fourier transform over the
additive group \(\mathbb{Z}_3^N\).

To see this, we expand the state in the computational basis as
\(\ket{\psi_{\mathbf a}}=\sum_{\mathbf x\in\mathbb{Z}_3^N}\alpha_{\mathbf x}\ket{\mathbf x}\).
Using \(A_{\mathbf 0}\ket{\mathbf x}=\ket{-\mathbf x}\) and
\(Z_{\mathbf b}\ket{\mathbf x}=\omega^{\mathbf b\cdot \mathbf x}\ket{\mathbf x}\),
a direct computation yields
\begin{equation}
\label{eq:mana_FT}
\chi_{\mathbf b}(\mathbf a)
\equiv
\bra{\psi_{\mathbf a}} Z_{\mathbf b} A_{\mathbf 0} Z_{\mathbf b}^\dagger \ket{\psi_{\mathbf a}}
=
\sum_{\mathbf x\in\mathbb{Z}_3^N}
\alpha_{\mathbf x}^*\,\alpha_{-\mathbf x}\,
\omega^{2\,\mathbf b\cdot \mathbf x},
\qquad \mathbf b\in\mathbb{Z}_3^N,
\end{equation}
with all arithmetic performed modulo \(3\).
Thus, for each fixed \(\mathbf a\), the full set of overlaps
\(\{\chi_{\mathbf b}(\mathbf a)\}_{\mathbf b\in\mathbb{Z}_3^N}\)
is obtained by applying a tensor-product Fourier transform to the vector
\(v_{\mathbf x}=\alpha_{\mathbf x}^*\alpha_{-\mathbf x}\).

To make the analogy with the qubit relation~\eqref{eq:Z_ovN_simplified}
explicit, introduce the unnormalized \(3\times 3\) ``Hadamard'' matrix
\begin{equation}
\label{eq:F3_unnorm_def}
(F_3)_{j,k} \;=\; \omega^{2 j k},
\qquad j,k\in\{0,1,2\},
\end{equation}
and consider its tensor power \(F_3^{\otimes N}\) acting on vectors indexed by
\(\mathbb{Z}_3^N\).
Then Eq.~\eqref{eq:mana_FT} can be written simply as
\begin{equation}
\label{eq:z3_FT_tensor_unnorm}
\bigl(\chi_{\mathbf b}(\mathbf a)\bigr)_{\mathbf b\in\mathbb{Z}_3^N}
\;=\;
F_3^{\otimes N}\,\bigl(v_{\mathbf x}\bigr)_{\mathbf x\in\mathbb{Z}_3^N},
\end{equation}
which is the direct qutrit analogue of the qubit formula for SRE~\eqref{eq:Z_ovN_simplified} with \(H^{\otimes N}\)
replaced by \(F_3^{\otimes N}\).

\begin{Algorithm}{Efficient calculation of mana using a fast \(\mathbb{Z}_3^N\) Fourier transform.
\label{alg:mana_fht}\vspace{-0.2cm}}
\Require Number of qutrits \(N\); state \(\ket{\psi}\).
\State \(m \gets 0\), \(\mathbf a\gets (0,\ldots,0)\), \(\ket{\psi_{\mathbf a}}\gets \ket{\psi}\).
\For{\(k=0\) to \(3^N-1\)} \Comment{Loop over \(\mathbf a\in\mathbb{Z}_3^N\).}
  \State Extract amplitudes \(\alpha_{\mathbf x}\) of \(\ket{\psi_{\mathbf a}}=\sum_{\mathbf x}\alpha_{\mathbf x}\ket{\mathbf x}\).
  \State Form \(v_{\mathbf x}\gets \alpha_{\mathbf x}^*\,\alpha_{-\mathbf x}\) for all \(\mathbf x\in\mathbb{Z}_3^N\).
  \State $(\chi_{\mathbf b})_{\mathbf b} \gets \textsc{FastFT}_{\mathbb{Z}_3^N}\bigl((v_{\mathbf x})_{\mathbf x}\bigr)$
  \Comment{Compute Eq.~\eqref{eq:mana_FT}.}
  \State \(m \gets m + \sum_{\mathbf b\in\mathbb{Z}_3^N} \bigl|\chi_{\mathbf b}\bigr|\)
  \State Update \(\mathbf a \gets \textsc{NextGray}_3(\mathbf a)\); update \(\ket{\psi_{\mathbf a}}\) accordingly.
\EndFor
\State \Return \(\mathcal{M}(\ket{\psi}) = \log_2 (m)\)
\vspace{-0.2cm}
\end{Algorithm}

Algorithm~\ref{alg:mana_fht} implements this strategy by replacing the
explicit sweep over all phase-space point operators with a fast
\(\mathbb{Z}_3^N\) Fourier transform for each fixed \(\mathbf a\).
In this way, the cost of the inner loop is reduced from \(O(9^N)\) to
\(O(N3^N)\), while the outer loop over \(\mathbf a\) still involves
\(O(3^N)\) iterations.
The resulting time complexity is \(O(N9^N)\), providing an exponential
speedup over the naive \(O(27^N)\) approach of Algorithm~\ref{alg:mana_gray}.
The memory footprint remains \(O(3^N)\), set by storing the state vector
\(\ket{\psi_{\mathbf a}}\) and work arrays of length \(3^N\).

\subsection{Mana of mixed states via a fast tensor-product transform}
\label{sec:mana_mixed_fast}

When the target state is specified by a density matrix \(\rho\) (possibly
mixed), the computational strategy must be modified, since expectation
values can no longer be reduced to state-vector overlaps.
For qubits, the stabilizer R\'enyi entropy is a magic monotone \emph{only}
for pure-state stabilizer protocols~\cite{leone2024stabilizer}.
Nevertheless, the SRE can be used to construct a magic \emph{witness} for
mixed states~\cite{Haug25witnessing}.
To compute the SRE for a mixed state \(\rho\)~\cite{tarabunga23gauge} of
\(N\) qubits, one can proceed by calculating the full set of Pauli
expectation values \(\Tr(P_j\rho)\), as summarized in Appendix~B.1 of
Ref.~\cite{Hamaguchi24handbook}.
This yields an in-place Pauli-decomposition algorithm with runtime
\(O(N4^N)\) and memory cost \(O(4^N)\), improving over the naive
\(O(8^N)\) sweep.
Closely related routines for Pauli decomposition of dense operators have
also been developed in Refs.~\cite{Hantzko2024Tensorize,
Jones2024Decomposing, VidalRomero2023}.

In contrast to SRE, mana in qutrit systems remains a faithful non-stabilizerness monotone for mixed states~\cite{Veitch2014theresourcetheory}, making its calculation for mixed states particularly relevant. Accordingly, below we present an exact fast-transform algorithm that efficiently evaluates the full set of phase-space expectation values \(\Tr(\rho A_{\mathbf u})\) and thereby enables the computation of mana for mixed states.

For an \(N\)-qutrit density matrix \(\rho\), mana  $\mathcal{M}(\rho)$ is obtained from the discrete Wigner function
\(W_{\mathbf u}=\Tr(\rho A_{\mathbf u})\) by summing \(|W_{\mathbf u}|\) over all \(9^N\) phase-space points $\mathbf u\in \mathbb{Z}_3^{2N}$, cf.~Eq.~\eqref{eq:mana}. 
In contrast to the pure-state setting, where overlaps \(\bra{\psi}A_{\mathbf u}\ket{\psi}\) can be generated from a state vector, in the mixed-state setting one must evaluate \(\Tr(\rho A_{\mathbf u})\) for all \(\mathbf u\in\mathbb{Z}_3^{2N}\). 
A naive calculation of the $9^N$ traces yields a computational cost $O(3^{3N})$.
The key observation enabling an exponential reduction of the computational cost is that the entire collection \(\{ \Tr(\rho A_{\mathbf u})\}_{\mathbf u \in \mathbb{Z}^{2N}_3}\) can be written as a single linear map acting on a vectorized density matrix, and, crucially, this map factorizes as a tensor power of a fixed \(9\times 9\) single-qutrit matrix. This structure enables an exact fast tensor-product transform implementation of $\mathcal{M}(\rho)$ calculation that utilizes \(M^{\otimes N}\) without ever forming a \(9^N\times 9^N\) matrix.

\subsubsection{Single-qutrit construction.}

We collect the expectation values of the phase-space point operators in a vector
$w\in\mathbb{C}^9$ with components
\begin{equation}
w_u \;=\; \mathrm{Tr}(\rho A_u), \qquad u\in\mathbb{Z}_3^2,
\end{equation}
and represent the density matrix $\rho$ in the column-stacking convention
\begin{equation}
\mathrm{vec}(\rho)
=
\bigl(\rho_{00},\rho_{10},\rho_{20},
      \rho_{01},\rho_{11},\rho_{21},
      \rho_{02},\rho_{12},\rho_{22}\bigr)^{\mathsf{T}}
\in\mathbb{C}^9.
\end{equation}
Using $\mathrm{Tr}(\rho A_u)
= \mathrm{vec}(A_u^{\mathsf{T}})^{\mathsf{T}}\mathrm{vec}(\rho)$, we can write
\begin{equation}
w \;=\; M\,\mathrm{vec}(\rho),
\end{equation}
where the $9\times 9$ matrix $M$ is defined as
\begin{equation}
M_{u,\alpha} \;=\; \bigl[\mathrm{vec}(A_u^{\mathsf{T}})\bigr]_\alpha,
\qquad
u\in\mathbb{Z}_3^2,\ \alpha\in\{0,\dots,8\}.
\end{equation}

The matrix $M$ has a particularly simple structure.
In the $(u=(a,a'))$ row and $(q,p)$ column labelling induced by the
vectorization above, each row of $M$ has exactly three non-zero entries,
with values in $\{1,\omega,\omega^2\}$.
More precisely, for fixed $a'$ the three rows with $a=0,1,2$ have support
on a common ``line'' of three operator-basis elements $(q,p)$ satisfying
$q+p\equiv -a'\ (\mathrm{mod}\;3)$, and on that line the $3\times 3$
submatrix is (up to permutations) the qutrit matrix  $F_3$ given by Eq.~\eqref{eq:F3_unnorm_def}. Accordingly, there exist permutation matrices $P_{\mathrm{row}}$ and
$P_{\mathrm{col}}$ such that
\begin{equation}
M
\;=\;
P_{\mathrm{row}}^{\mathsf{T}}\,
\bigl(F_3 \oplus F_3 \oplus F_3\bigr)\,
P_{\mathrm{col}},
\end{equation}
i.e.\ in a suitably ordered basis the phase-space transform on a single
qutrit decomposes as a direct sum of three identical $3\times 3$ Fourier
blocks.
%In particular, $M M^\dagger = 3\,\mathbb{1}_9$, so that $M/\sqrt{3}$ is unitary.

\subsubsection{Construction for many qutrits.}
\label{subsec:mana_mixed_fast_sub}
Let \(\rho\) be a density operator for system of $N$ qutrits, with computational basis
\(\{\ket{\mathbf{i}}\}_{\mathbf{i}\in\mathbb{Z}_3^N}\), \(\mathbf{i}=(i_0,\ldots,i_{N-1})\).
We write matrix elements as
\(\rho_{\mathbf{i},\mathbf{j}} = \langle \mathbf{i} | \rho | \mathbf{j} \rangle\),
and regard \(\rho\) as a tensor with indices \((i_0,\ldots,i_{N-1};j_0,\ldots,j_{N-1})\).
At each site \(k\), we combine the bra/ket indices into a single local label
\begin{equation}
  \mu_k \;\equiv\; 3 i_k + j_k \in \{0,\ldots,8\}.
\end{equation}
This identifies the local operator space with \(\mathbb{C}^9\) spanned by operators \(\{\ket{\mu_k}\}_{\mu_k=0}^8\), and we define
$\mathrm{Vec}_N$ as mapping from the operator space $\mathcal{L}\big((\mathbb{C}^3)^{\otimes N}\big)$ to vector space $(\mathbb{C}^9)^{\otimes N}$ by
\begin{equation}
  \ket{\mathbf{i}}\!\bra{\mathbf{j}}
  \;\longmapsto\;
  \ket{\mu_0}\otimes\cdots\otimes\ket{\mu_{N-1}} .
\end{equation}
Equivalently, the resulting vector \(v = \mathrm{Vec}_N(\rho)\in\mathbb{C}^{9^N}\) is indexed by the
base-9 integer
\begin{equation}
  p(\mathbf{i},\mathbf{j})
  = \sum_{k=0}^{N-1} \mu_k\,9^{N-1-k},
\end{equation}
so that \(v_{p(\mathbf{i},\mathbf{j})} = \rho_{\mathbf{i},\mathbf{j}}\).
This convention ensures that product operators factorize as
\(\mathrm{Vec}_N(B^{(1)}\otimes\cdots\otimes B^{(N)})
 = \mathrm{vec}(B^{(1)})\otimes\cdots\otimes\mathrm{vec}(B^{(N)})\).

For any phase-space point operator \(A_{\mathbf u}=A_{u_0}\otimes\cdots\otimes A_{u_{N-1}}\) with \(\mathbf u\in\mathbb{Z}_3^{2N}\), the vector of the expectation values
\(w_{\mathbf u}=\Tr(\rho A_{\mathbf u})\), ordered consistently with the base-9 encoding of \(\mathbf u\),  can be calculated as
\begin{equation}
\label{eq:MN_factor_mana}
w = M^{\otimes N}\,\mathrm{Vec}_N(\rho),
\end{equation}
because both the operator basis and the vectorization respect tensor products. Thus, computing all \(9^N\) expectation values reduces to applying \(M^{\otimes N}\) to a length-\(9^N\) vector, in an analogy to the structure of SRE and mana computation for pure states unravelled by Eq.~\eqref{eq:Z_ovN_simplified} and \eqref{eq:z3_FT_tensor_unnorm}, respectively.

To efficiently implement Eq.~\eqref{eq:MN_factor_mana}, in Algorithm~\ref{alg:mana_mixed_fast_detailed}, we apply \(M^{\otimes N}\) \emph{in place} by sweeping over
the tensor indices (``legs''), each corresponding to one qutrit. We regard the vector \(v\in\mathbb{C}^{9^N}\) as an
\(N\)-index array with local dimension \(9\), i.e.\ as a tensor \(v_{\mu_0,\ldots,\mu_{N-1}}\). For a fixed leg
\(\ell=0,\ldots,N-1\), we hold all other indices \(\mu_{k\neq\ell}\) fixed and apply the single-qutrit matrix \(M\)
along the remaining index \(\mu_{\ell}\). The corresponding set of nine entries
\(\{v_{\mu_0,\ldots,\mu_{\ell},\ldots,\mu_{N-1}}\}_{\mu_{\ell}=0}^8\) is what we refer to as a \emph{fiber} along leg
\(\ell\).
In our memory layout, these fibers appear as nine equally spaced elements in the flat array. Writing indices in base~9,
entries that differ only in the \(\ell\)-th digit are separated by $ n_s = 9^{\,N-1-\ell} $
in memory. Each contiguous block of length \(9 n_s\) contains all fibers obtained by varying \(\mu_{\ell}\) while holding
the remaining indices fixed. For every block (with starting index \(\mathrm{base}\)) and every offset
\(m\in\{0,\ldots,n_s-1\}\), we gather the nine elements
$ 
  \bigl\{ v[\mathrm{base} + m + a\,n_s] \bigr\}_{a=0}^8
$
into a local vector \(x\), compute \(y = M x\), and write the result back to the corresponding locations. Sweeping over all
offsets \(m\) and all blocks in this way processes every fiber exactly once for each \(\ell\), and thus realizes the
full action of \(M^{\otimes N}\) without ever forming the \(9^N\times 9^N\) matrix explicitly.

\begin{Algorithm}{Exact computation of mixed-state mana via application of \(M^{\otimes N}\).
\label{alg:mana_mixed_fast_detailed}\vspace{-0.2cm}}
\Require Number of qutrits \(N\); density matrix \(\rho\in\mathbb{C}^{3^N\times 3^N}\) stored column-major.
\State Construct \(M\in\mathbb{C}^{9\times9}\) row-wise via \(M_{u,:}=\vec(A_u^{\mathsf T})^{\mathsf T}\) (column-major vec).
\State Compute \(v=\mathrm{Vec}_N(\rho)\in\mathbb{C}^{9^N}\) as follows:
\For{\(c=0\) to \(3^N-1\)} \Comment{column index of \(\rho\)}
  \State Decode \(c\) into base-3 digits \(\mathbf j=(j_0,\ldots,j_{N-1})\).
  \For{\(r=0\) to \(3^N-1\)} \Comment{row index of \(\rho\)}
    \State Decode \(r\) into base-3 digits \(\mathbf i=(i_0,\ldots,i_{N-1})\).
    \State Set \(\mu_k=3i_k+j_k\) and initialize \(p\leftarrow 0\); then update \(p\leftarrow 9p+\mu_k\) for \(k=0,\ldots,N-1\).
    \State Assign \(v[p]\gets \rho_{r,c}\) (column-major: \(\rho_{r,c}=\rho[r+c\,3^N]\)).
  \EndFor
\EndFor
\State \textbf{Apply \(M^{\otimes N}\) in place (leg sweep):}
\For{\(\ell=0\) to \(N-1\)} \Comment{\(\ell=0\) is the most-significant base-9 digit}
  \State \(n_s \gets 9^{\,N-1-\ell}\) ,  \(b_s \gets 9\,n_s\), \(n_b \gets 9^N / b_s\)
  \For{\(b=0\) to \(n_b-1\)}
    \State \(\mathrm{base}\gets b\cdot b_s\).
    \For{\(m=0\) to \(n_s-1\)} \Comment{\(m\) labels fibers within the block}
      \State Gather the fiber along leg \(\ell\): \quad \(x_a\gets v[\mathrm{base}+m+a\,n_s]\) for \(a=0,\ldots,8\).
      \State Compute \(y\gets Mx\) (dense \(9\times9\) matrix-vector product).
      \State Write back \(v[\mathrm{base}+m+a\,n_s]\gets y_a\) for \(a=0,\ldots,8\).
    \EndFor
  \EndFor
\EndFor
\State Now \(v = w\) with entries \(w_{\mathbf u}=\Tr(\rho A_{\mathbf u})\) in the base-9 ordering of \(\mathbf u\).
\State \Return \(\mathcal{M}(\rho) = \log_2 \sum_{\mathbf u} | w_\mathbf{u}|\) consistently with Eq.~\eqref{eq:mana}.
\vspace{-0.2cm}
\end{Algorithm}

The cost per leg is the number of fibers (\(9^{N-1}\)) times the constant cost of a dense \(9\times 9\) matrix-vector multiplication, giving \(O(9^{N})\) operations per leg. This results in a total computational  complexity \(O(N\,9^{N})\) of the  Algorithm~\ref{alg:mana_mixed_fast_detailed} . The memory footprint is \(O(9^N)\), dominated by storing the density matrix \(v =  \mathrm{Vec}_N(\rho) \).

\section{Examples and benchmarks}
\label{subsec:uses}

In this section we benchmark Algorithms~\ref{alg:sre_eff}, \ref{alg:sre_sampling_fht},
\ref{alg:mana_fht}, and~\ref{alg:mana_mixed_fast_detailed}, as implemented in
\href{https://github.com/bsc-quantic/HadaMAG.jl/}{HadaMAG.jl}
, on states generated
by random quantum circuits. Such circuits provide a canonical source of highly
entangled many-body states, for which state-vector methods are the natural tools
of choice. We consider a one-dimensional chain of $N$ qubits ($d=2$) or qutrits ($d=3$),
evolved by a brick-wall circuit. The evolution operator is
$U_t = \prod_{r=1}^t U^{(r)}$, where $t$ is the circuit depth (or time), with
layers
\begin{equation}
U^{(2m)} = \prod_{i=1}^{N/2-1} u_{2i,2i+1};,\quad
U^{(2m+1)} = \prod_{i=1}^{N/2} u_{2i-1,2i};,
\label{eq:randC}
\end{equation}
where each two-qudit gate $u_{i,j}$ is drawn from the unitary group
$\mathcal{U}(d^2)$ with respect to the Haar measure~\cite{mezzadri2006generate}.

The SRE and mana routines in
\href{https://github.com/bsc-quantic/HadaMAG.jl/}{HadaMAG.jl}
 support several
backends---\texttt{:serial}, \texttt{:threads}, \texttt{:mpi\_threads},
\texttt{:cuda}, and \texttt{:cuda\_threads}---which exploit different levels of
parallelism or support GPU acceleration, as illustrated in the examples below and detailed in  \href{https://bsc-quantic.github.io/HadaMAG.jl/dev/}{the documentation}.

\subsection{Exact calculation of stabilizer R\'{e}nyi entropy}
\label{subsec:exactSRE}

We begin by benchmarking the numerically exact computation of SRE for qubits.
Starting from the product state $\ket{0}^{\otimes N}$, we generate a Haar-random
brick-wall circuit state by applying Eq.~\eqref{eq:randC}, implemented in
\texttt{HadaMAG.jl} as
\texttt{psi = rand\_haar(N; depth = t)}.
We then evaluate the stabilizer R\'enyi entropy $M_2(\ket{\psi})$ with
Algorithm~\ref{alg:sre_eff} using the call \texttt{SRE(psi, 2)}.

        \begin{juliabox}
julia> using HadaMAG

julia> psi = rand_haar(16; depth=0)
StateVec{ComplexF64,2}(n=16, dim=65536, mem=1.0 MiB)

julia> m,l=SRE(psi, 2); println("SRE=\$m  lost_norm=\$l")
SRE=-0.0  lost_norm=0.0)

julia> psi=rand_haar(16; depth=1); m,l=SRE(psi, 2); println("SRE=\$m  lost_norm=\$l")
SRE=5.519381245826574 lost_norm=2.220446049250313e-15

julia> psi=rand_haar(16; depth=16); m,l=SRE(psi, 2);println("SRE=\$m  lost_norm=\$l")
SRE=13.985447072668483 lost_norm=3.3306690738754696e-15
        \end{juliabox}

The initial state $\ket{0}^{\otimes N}$ is a stabilizer state and, accordingly,
we find $M_2(\ket{0}^{\otimes N}) = 0$.
As the circuit depth $t$ increases, the SRE grows monotonically and, for large
$t$, approaches the value characteristic of Haar-random states,
$M^{\mathrm{Haar}}_2 = \log_2(2^N+3)-2$~\cite{Turkeshi23flat}.
In the dynamics of random circuits, this saturation to the Haar value occurs at
circuit depths that scale only logarithmically with the system size $N$, as
analyzed quantitatively in Ref.~\cite{turkeshi2024magic}. The second number \texttt{lost\_norm} returned by \texttt{SRE} routine, is equal to $1- \sum_{P\in\mathcal{P}_N} \braket{\psi|P|\psi}^2 / 2^N $, which for a normalized state $\ket{\psi}$ should be numerically equal to $0$, cf., Eq.~\eqref{eq:distNAT}.
To exploit parallelism in
\href{https://github.com/bsc-quantic/HadaMAG.jl/}{HadaMAG.jl}, it suffices to
select an appropriate backend in the SRE routine. For instance, multithreading
within a single computing node is enabled by
\begin{juliabox}
julia> using HadaMAG

julia> psi = rand_haar(16; depth=4)
StateVec{ComplexF64,2}(n=16, dim=65536, mem=1.0 MiB)

julia> m,l = SRE(psi, 2; backend=:threads); println("SRE=\$m  lost_norm=\$l")
SRE=11.120916275249016 lost_norm=2.9976021664879227e-15
\end{juliabox}
\noindent
where the number of threads is controlled by the environment variable
\texttt{JULIA\_NUM\_THREADS}.
In Algorithm~\ref{alg:sre_eff}, parallelization is achieved by partitioning the
loop over the binary Gray code into chunks, each associated with a distinct set
of $X$-strings. Since these chunks are independent, the algorithm exhibits a
high degree of parallelism and yields nearly ideal speedup as the number
$n_{\mathrm{cores}}$ of cores is increased.

The performance of
\href{https://github.com/bsc-quantic/HadaMAG.jl/}{HadaMAG.jl} on a single node
with $n_{\mathrm{cores}}=112$ is summarized in Table~\ref{tabXXZ}, and allows
one to compute the SRE for systems up to $N=22$ qubits in under one hour.
For comparison, an implementation of the naive
Algorithm~\ref{alg:sregray} is limited to $N=16$ qubits within the same wall
time on a single node, illustrating the exponential speedup provided by the
fast Hadamard transform.

The package \href{https://github.com/bsc-quantic/HadaMAG.jl/}{HadaMAG.jl}
also supports MPI-based parallelism via the \texttt{:mpi\_threads} backend;
see \href{https://bsc-quantic.github.io/HadaMAG.jl/dev/}{the documentation}
for details.
In addition, we provide efficient CUDA kernels for
Algorithm~\ref{alg:sre_eff}, accessible through the \texttt{:cuda} and
\texttt{:mpi\_cuda} backends.
For example, computing the SRE with
\href{https://github.com/bsc-quantic/HadaMAG.jl/}{HadaMAG.jl} on multiple
GPUs with MPI communication can be done as follows:
\begin{juliabox}
using HadaMAG
using CUDA
using MPI
using Random

MPI.Initialized() || MPI.Init()
comm = MPI.COMM_WORLD; rank = MPI.Comm_rank(comm)

Random.seed!(123); psi = rand_haar(L; depth=5)

batch_size = 128; nthreads_per_device = 128
m2, lost_norm = SRE(psi, 2; backend=:mpi_cuda, progress=false,batch=batch_size,threads=nthreads_per_device)
\end{juliabox}

Table~\ref{tabXXZ} summarizes benchmarks for the MPI-based backends on
\(n_{\mathrm{nodes}} = 8\) compute nodes, each equipped with \(112\) CPU
cores, and compares them with SRE computations performed on \(8\) nodes
where each node uses \(4\) GPUs.
Offloading the core kernels to GPUs yields a substantial performance gain
for system sizes \(N \gtrsim 18\) qubits, allowing us to compute the SRE
for \(N=25\) qubits in roughly \(2\) hours of wall-clock time.

We note that the reported memory usage is significantly larger than the
memory required to store the state vector and a single set of auxiliary
arrays: each thread needs its own auxiliary workspace for the fast Hadamard
transform, so the total memory consumption grows with the number of
threads.
Although alternative designs are possible, in
\href{https://github.com/bsc-quantic/HadaMAG.jl/}{HadaMAG.jl} we have
adopted this choice to minimize total runtime; for the system sizes of
interest, the resulting memory footprint remains well within the
capabilities of current computing architectures.

\begin{table}[h]
\centering
\begin{minipage}[t]{0.45\linewidth}
\centering
{\setlength{\tabcolsep}{3pt}
\begin{tabular}{@{}c@{\hspace{0.2cm}}@{}c@{} c c c c c}
\toprule
 &  & {$N$}
   & {$t_w$}
   & {$t_{\text{CPU}}^{\text{tot}}$}
   & {$n_{\text{cores}}$}
   & {RAM [MB]} \\
\midrule
\multirow{3}{*}{\rotatebox[origin=c]{90}{naive}} 
  &  & 12  & $0.73\,\text{s}$     & $1.37\,\text{min}$ & 112 & 0.13 \\
  &  & 14  & $41.4\,\text{s}$     & $1.29\,\text{h}$   & 112 & 0.32 \\
  &  & 16  & $44.2\,\text{min}$   & $82.5\,\text{h}$   & 112 & 1.07 \\
\midrule
\multirow{6}{*}{\rotatebox[origin=c]{90}{\href{https://github.com/bsc-quantic/HadaMAG.jl/}{HadaMAG.jl}}} 
  &  & 12  & $18\,\text{ms}$      & $2.02\,\text{s}$   & 112 & 5 \\
  &  & 14  & $32\,\text{ms}$      & $3.63\,\text{s}$   & 112 & 17 \\
  &  & 16  & $0.32\,\text{s}$     & $35.9\,\text{s}$   & 112 & 68 \\
  &  & 18  & $4.83\,\text{s}$     & $9.01\,\text{min}$ & 112 & 269 \\
  &  & 20  & $2.66\,\text{min}$   & $4.96\,\text{h}$   & 112 & 1074 \\
  &  & 22  & $55.0\,\text{min}$   & $103\,\text{h}$    & 112 & 4295 \\
\bottomrule
\end{tabular}}
\end{minipage}%
\hspace{0.5cm}
\begin{minipage}[t]{0.45\linewidth}
\centering
{\setlength{\tabcolsep}{3pt}
\begin{tabular}{@{}c@{\hspace{0.2cm}}@{}c@{} c c c c c}
\toprule
 &  & {$N$}
   & {$n_{\text{nodes}}$}
   & {$t_w$}
   & {$t^{\text{tot}}$}
   & {RAM [MB]} \\
\midrule
\multirow{5}{*}{\rotatebox[origin=c]{90}{SRE (GPU)}} 
  &  & 18 & 8 & $0.78\,\text{s}$     & $24.96\,\text{s}$      & 30 \\
  &  & 20 & 8 & $8.23\,\text{s}$     & $4.39\,\text{min}$     & 120 \\
  &  & 22 & 8 & $1.87\,\text{min}$   & $59.84\,\text{min}$    & 481 \\
  &  & 24 & 8 & $0.06\,\text{min}$   & $1.92\,\text{min}$     & 1929 \\
  &  & 25 & 8 & $2.10\,\text{h}$     & $67.20\,\text{h}$      & 3866 \\
\midrule
\multirow{5}{*}{\rotatebox[origin=c]{90}{SRE (CPU)}} 
  &  & 18 & 8 & $0.76\,\text{s}$     & $11.35\,\text{min}$    & 43 \\
  &  & 20 & 8 & $21.46\,\text{s}$    & $5.34\,\text{h}$       & 174 \\
  &  & 22 & 8 & $6.88\,\text{min}$   & $102.74\,\text{h}$     & 694 \\
  &  & 24 & 8 & $2.24\,\text{h}$     & $2007\,\text{h}$    & 2778 \\
  &  & 25 & 8 & $9.44\,\text{h}$     & $8458\,\text{h}$    & 5556 \\
\bottomrule
\end{tabular}}
\end{minipage}
\caption{\label{tabXXZ}
Performance of SRE computation in \href{https://github.com/bsc-quantic/HadaMAG.jl/}{HadaMAG.jl}.
\textbf{Left:} CPU-only runs, comparing a custom implementation of Algorithm~\ref{alg:sregray}
(``naive'') with Algorithm~\ref{alg:sre_eff} implemented in
\href{https://github.com/bsc-quantic/HadaMAG.jl/}{\text{HadaMAG.jl}}.
For each system size $N$, we report the total wall-clock time $t_w$, the
aggregate CPU time $t_{\text{CPU}}^{\text{tot}}$ summed over all
$n_{\text{cores}}=112$ cores of a single node, and the peak memory usage.
The CPU used: Intel Sapphire Rapid
8480+, 112 cores/node).
\textbf{Right:} Comparison between the CPU and GPU backends of
\href{https://github.com/bsc-quantic/HadaMAG.jl/}{HadaMAG.jl} for $N=18$–$25$.
Each job uses $n_{\text{nodes}}=8$ nodes, corresponding to $8\times 112 $ CPU cores and $8\times4$ GPUs, respectively; we report the wall-clock time $t_w$,
the total device time $t^{\text{tot}}$ (CPU-core hours for the CPU runs,
GPU-device hours for the GPU runs), and peak RAM. The GPUs used: Nvidia Hopper with 64 HBM2.
}
\end{table}

\subsection{Sampling algorithm for qubit SRE}
The \(O(N4^N)\) scaling of Algorithm~\ref{alg:sre_eff}, reflected in the
total CPU/GPU times reported in Table~\ref{tabXXZ}, implies that the
computational cost of \emph{exact} SRE evaluation reaches thousands of CPU
hours already for \(N=24\).
By contrast, applying a single two-qubit gate from the circuit
Eq.~\eqref{eq:randC} costs only \(O(2^N)\), so computing time-evolved state
vectors of size \(N \approx 30\) qubits is well within reach on
contemporary hardware.
In this regime, Algorithm~\ref{alg:sre_sampling_fht} becomes a natural
method of choice for \emph{approximate} SRE estimation: its cost scales as
\(O\bigl(n_S\,N2^N\bigr)\), where \(n_S\) is the number of Monte Carlo
samples.
Below, we compare the numerically exact SRE for \(N=16\) qubits with the
estimate obtained from Algorithm~\ref{alg:sre_sampling_fht} using
\(n_S = 1000\).

        \begin{juliabox}
julia> using HadaMAG

julia> L = 16; psi = rand_haar(L; depth=2)
StateVec{ComplexF64,2}(n=16, dim=65536, mem=1.0 MiB)

julia> m,l=SRE(psi, 2); println("SRE=\$m  lost_norm=\$l")
SRE=14.00006877485478  lost_norm=2.460254222569347e-13)

julia> m_approx=MC_SRE(psi, 2; Nsamples=1000); println("SRE_approx=\$m_approx")
SRE_approx=14.000013644355313
        \end{juliabox}
        
\begin{figure*}[h]
    \centering
    \includegraphics[width=1\linewidth]{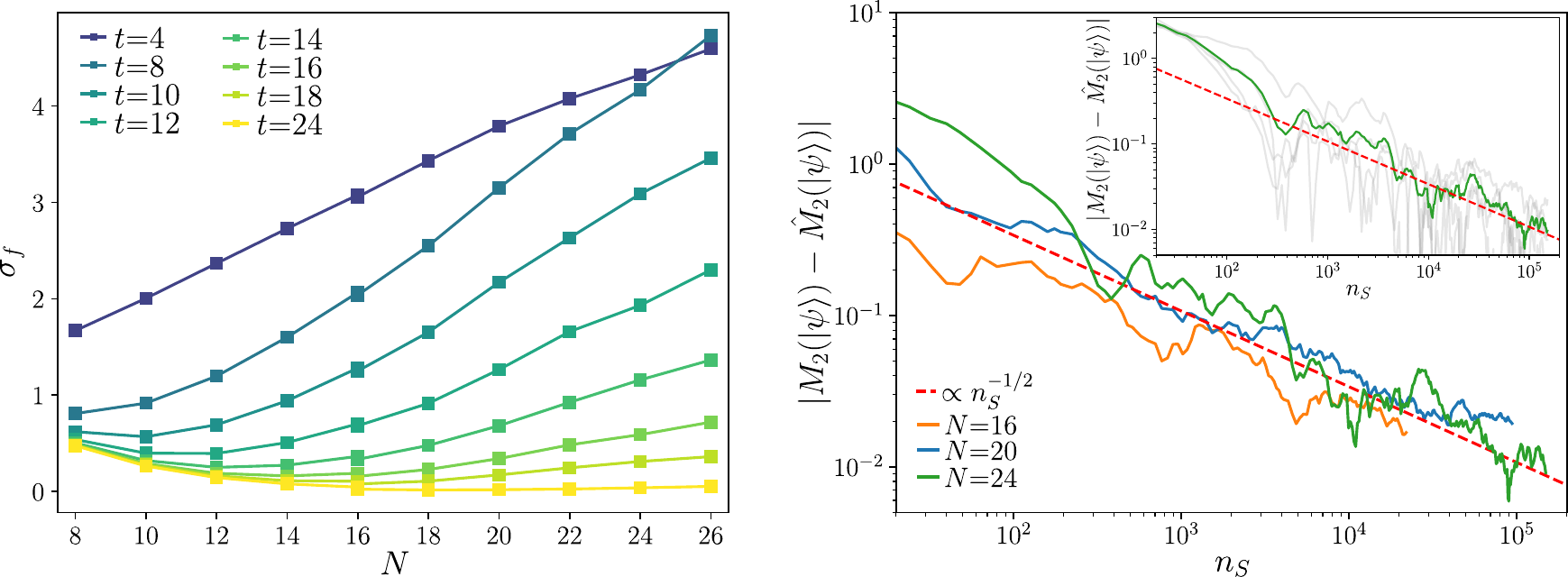}
    \caption{
\textbf{Left:} Variance \(\sigma_f\) of the ``energy'' \(f(X_{\mathbf a})\) at \(\beta=1\),
which controls the statistical uncertainty of the SRE in the sampling
Algorithm~\ref{alg:sre_sampling_fht}, for states evolved under quantum
circuits~\eqref{eq:randC} of depth \(t\), plotted as a function of system size \(N\).
\textbf{Right:} Absolute error of the SRE \(M_2\) at circuit depth \(t=4\),
comparing the numerically exact result from Algorithm~\ref{alg:sre_eff} with its
sampling estimate \(\hat{M}_2\), as a function of the number of samples \(n_S\).
The data are averaged over four independent realizations of the sampling
procedure. The inset shows the absolute SRE error (in grey) for each of the four
realizations and their average for \(N=24\).
Consistently with Eq.~\eqref{eq:m2error}, the absolute error decreases as
\(n_S^{-1/2}\), as indicated by the red dashed lines.
    }
    \label{fig:fide}
\end{figure*}

In the example above, the difference between the exact value of the SRE and
its Monte Carlo estimate is approximately \(5.5\times 10^{-5}\).
This small discrepancy reflects the fact that the variance \(\sigma_f^2\) of
the ``energy'' \(f(X_{\mathbf a})\) appearing in
Eq.~\eqref{eq:feP}---to which the statistical error \(\delta M_2\) is
proportional, cf.\ Eq.~\eqref{eq:m2error}---is particularly small for states
close to Haar-random, such as those generated at circuit depths
\(t \approx N\).

The variance \(\sigma_f^2\), which sets the number of samples \(n_S\) required
to reach a given precision in the SRE estimate, depends on the state under
consideration. In Fig.~\ref{fig:fide} we analyze its behavior for states
generated by random quantum circuits. We find that, beyond a $t$-dependent threshold system
size, \(\sigma_f\) is well approximated by a linear function of \(N\). At fixed
\(N\), the variance decreases monotonically with increasing circuit depth
\(t\)~\footnote{We exclude the identity Pauli string \(P=\mathbb{I}\) from the
sampling procedure and account explicitly for its contribution, equal to unity
for any \(\ket{\psi}\), in the post-processing of the results,
cf.~\cite{Turkeshi25spectrum}.}.
This linear growth of \(\sigma_f\) translates into a relatively mild, quadratic
increase in the number of samples \(n_S\) needed to maintain a fixed
statistical error on the SRE estimate as \(N\) grows. The right panel of
Fig.~\ref{fig:fide} illustrates the convergence of the sampling algorithm
towards the numerically exact value of \(M_2\) as the sample number \(n_S\) is
increased.

\subsection{Mana calculation}

Computation of mana in \href{https://github.com/bsc-quantic/HadaMAG.jl/}{HadaMAG.jl} follows a pattern similar to that used for the SRE. We begin by benchmarking the numerically exact evaluation of mana for qutrits using Algorithm~\ref{alg:mana_fht}. Starting from the product state $\ket{0}^{\otimes N}$, we generate a Haar-random brick-wall circuit state by applying Eq.~\eqref{eq:randC}, implemented for qutrits as \texttt{psi = rand\_haar(N; depth = t, q = 3)}, and then compute its mana $\mathcal{M}(\ket{\psi})$:

        \begin{juliabox}
julia> using HadaMAG

julia> psi = rand_haar(10; depth=0, q=3)
StateVec{ComplexF64,3}(n=10, dim=59049, mem=923.0 KiB)

julia> Mana(psi)
[==================================================] 100.0%  (59049/59049)
-1.9706086672703374e-14

julia> psi = rand_haar(10; depth=4, q=3)
StateVec{ComplexF64,3}(n=10, dim=59049, mem=923.0 KiB)

julia> Mana(psi)
[==================================================] 100.0%  (59049/59049)
4.77071931195802

        \end{juliabox}

As for the SRE, \href{https://github.com/bsc-quantic/HadaMAG.jl/}{\text{HadaMAG.jl}} provides several backends for mana computation, including \texttt{:serial}, \texttt{:threads}, and \texttt{:mpi\_threads}; see \href{https://bsc-quantic.github.io/HadaMAG.jl/dev/}{the documentation} for details. Using the MPI backend, we compute the mana of a pure state of $N=14$ qutrits in roughly $2$ hours on four computing nodes, while computation for $N=15$ qutrits takes about $11$ hours on $8$ computing nodes, as reported in Table~\ref{manaresults}.

We then evaluate the reduced density matrix $\rho$ for a subsystem of
$N_A < N$ qutrits by tracing out the degrees of freedom outside $A$, and
compute the mana $\mathcal{M}(\rho)$ of the resulting mixed state using
Algorithm~\ref{alg:mana_mixed_fast_detailed}.

        \begin{juliabox}
julia> using HadaMAG

julia> psi = rand_haar(10; depth=3, q=3)
StateVec{ComplexF64,3}(n=10, dim=59049, mem=923.0 KiB)

julia> rho=reduced_density_matrix(psi, 8)
DensityMatrix{ComplexF64,3}(n=8, dim=6561, mem=657.0 MiB)

julia> Mana(rho)
5.138534117189299
        \end{juliabox}

The mixed-state algorithm, Algorithm~\ref{alg:mana_mixed_fast_detailed}, differs
from the SRE and pure-state mana routines in that it does not involve a loop over
$X$-strings, making its parallelization more involved. At present,
\href{https://github.com/bsc-quantic/HadaMAG.jl/}{\text{HadaMAG.jl}} provides only the
\texttt{:serial} backend for mixed-state mana, enabling computations on a single CPU core.
Nevertheless, thanks to the underlying fast-transform scheme,
Algorithm~\ref{alg:mana_mixed_fast_detailed} allows us to compute the mana of reduced
density matrices $\rho$ for subsystems up to $N_A = 10$ qutrits within a few minutes. This routine has been employed in studies of local dynamics and spreading of non-stabilizerness in random quantum dynamics~\cite{Aditya25spreading}.
The main limitation for larger $N_A$ is not the runtime but the memory required to
store the full density matrix $\rho$, whose dimension grows as $3^{N_A}\times 3^{N_A}$.

\begin{table}[h]
\centering
\begin{minipage}[t]{0.58\linewidth}
\centering
{\setlength{\tabcolsep}{4pt}
\begin{tabular}{@{}c@{\hspace{0.2cm}}@{}c@{} c c c c c}
\toprule
 &  & {$N$}
   & {$n_{\text{nodes}}$}
   & {$t_w$}
   & {$t_{\text{CPU}}^{\text{tot}}$}
   & {RAM [MB]} \\
\midrule
\multirow{6}{*}{\rotatebox[origin=c]{90}{pure states}}
 &  & 6  & 4 & $14\,\text{ms}$      & $6.5\,\text{s}$       & 229 \\
 &  & 8  & 4 & $18\,\text{ms}$      & $8.1\,\text{s}$       & 344 \\
 &  & 10 & 4 & $0.12\,\text{s}$     & $54.8\,\text{s}$      & 573 \\
 &  & 12 & 4 & $1.28\,\text{min}$   & $9.58\,\text{h}$      & 2408 \\
 &  & 14 & 4 & $2.29\,\text{h}$     & $1026\,\text{h}$      & 18579 \\
 &  & 15 & 8 & $11.00\,\text{h}$    & $9859\,\text{h}$      & 54058 \\
\bottomrule
\end{tabular}}
\end{minipage}%
\hfill
\begin{minipage}[t]{0.38\linewidth}
\centering
{\setlength{\tabcolsep}{4pt}
\begin{tabular}{@{}c@{\hspace{0.2cm}}@{}c@{} c c c}
\toprule
 &  & {$N_A$}
   & {$t_w$}
   & {RAM [MB]} \\
\midrule
\multirow{5}{*}{\rotatebox[origin=c]{90}{mixed-states}}
  &  & 2  & $<1\,\text{ms}$      & $<1$ \\
  &  & 4  & $<1\,\text{ms}$      & $<1$ \\
  &  & 6  & $29\,\text{ms}$      & 9 \\
  &  & 8  & $3.06\,\text{s}$     & 706 \\
  &  & 10 & $5.60\,\text{min}$   & 57128 \\
\bottomrule
\end{tabular}}
\end{minipage}
\caption{ \label{manaresults} 
Performance of mana computation in \href{https://github.com/bsc-quantic/HadaMAG.jl/}{HadaMAG.jl}.
\textbf{Left:}
Mana computation with Algorithm~\ref{alg:mana_fht} for an $N$-qutrit random state using $n_{\mathrm{nodes}}=4,8$ nodes, with one core per MPI task and $112$ CPU cores per node.
Here $t_w$ is the wall-clock time, $t_{\text{CPU}}^{\text{tot}}$ is the total CPU time
($t_{\text{CPU}}^{\text{tot}} = t_w \times 4 \times 112$), and RAM is the total memory usage in megabytes.
\textbf{Right:}
Runtime and memory usage for mixed-state mana calculations with Algorithm~\ref{alg:mana_mixed_fast_detailed} on a subsystem of size $N_A$ using a single CPU core. 
Here $t_w$ is the wall-clock time and RAM is the peak memory consumption in megabytes. The benchmarks were performed on the same hardware as in Table.~\ref{tabXXZ}.
}
\end{table}

\section{Conclusion}
\label{sec:conclusion}

\subsection{Summary}
In this work we introduced a family of algorithms for the efficient
evaluation of SRE and mana, applicable to both
pure and mixed many-body states.
At the core of our approach are the numerically exact
Algorithm~\ref{alg:sre_eff} for SRE and Algorithm~\ref{alg:mana_fht} for
mana, whose computational cost scales as \(O(N d^{2N})\) for local Hilbert space
dimension \(d\).
Our algorithms provide an exponential improvement over the naive schemes (Algorithms~\ref{alg:sregray}
and~\ref{alg:mana_gray}), which scale as \(O(d^{3N})\), and achieve this
speedup without introducing any additional memory overhead: the memory
footprint remains \(O(d^{N})\), set by the state vector representation of the analyzed quantum states and a small number of work arrays of the same length.

From a practical standpoint, the introduced algorithms exhibit a high
degree of parallelism and are well suited for modern high-performance
computing architectures.
All core routines are implemented in the open-source
\href{https://github.com/bsc-quantic/HadaMAG.jl/}{HadaMAG.jl} package, which leverages multi-threading, MPI-based
distributed memory parallelism, and GPU acceleration via CUDA.
This enables numerically exact calculations of SRE and mana for system
sizes that are significantly beyond the reach of naive implementations,
and makes resource-theoretic diagnostics readily available for large-scale
numerical studies of quantum dynamics.

On the conceptual side, the fast Hadamard transform (for SRE) and its
ternary analogue (for mana) reveal a unifying mathematical structure
underlying seemingly different non-stabilizerness measures.
In the qubit case, the efficient SRE algorithm recasts the full set of
Pauli overlaps into a sequence of Walsh--Hadamard transforms on
\(\mathbb{Z}_2^N\), while in the qutrit case the efficient mana algorithm
expresses Wigner function evaluations as fast Fourier transforms on
\(\mathbb{Z}_3^N\).
This common ``fast transform'' viewpoint elucidates why both quantities can
be computed in \(O(N d^{2N})\) time.
A detailed comparison with related prior uses of the fast Walsh--Hadamard transform in the context of non-stabilizerness and Pauli decomposition is provided in Appendix~\ref{app:fwht_comparison}.
%and suggests possibilities of generalizations to higher local dimensions and other non-stabilizerness measures, such as the generalized stabilizer entropies~\cite{turkeshi2024magic}.

The same fast-transform framework also underpins our sampling scheme for
SRE, Algorithm~\ref{alg:sre_sampling_fht}, which combines thermodynamic
integration with a fast-Hadamard-based evaluation of effective energies.
By sampling only over \(X\)-patterns and using the fast transform to
collapse the sum over \(Z\)-patterns and employing the thermodynamic integration, the algorithm may avoid the exponential
growth in the required number of samples with system size that plagues
naive Monte Carlo schemes.
Our results for quantum circuit states suggest that for generic states the statistical uncertainty on the SRE \(M_2\) grows at most polynomially with \(N\), so that a fixed target precision can be maintained with a polynomial increase of sampling effort, rather than an exponential one.

For mixed qutrit states, we further showed that the fast-transform strategy
extends to mana, leading to the mixed-state
Algorithm~\ref{alg:mana_mixed_fast_detailed}.
Here the role of the simple Hadamard transform is played by a structured
linear map \(M^{\otimes N}\) that connects the vectorized density matrix to
the vector of phase-space expectation values.
Although the single-qutrit map \(M\) has a more intricate internal
structure than a Fourier matrix, it can still be applied in an in-place,
tensor-factorized fashion, yielding an overall complexity
\(O(N 9^{N}) = O(N d^{2N})\) and an exponential speedup over the naive
\(O(d^{3N})\) approach based on explicit evaluation of expectation values of all phase-space point operators.

Table~\ref{tab:complexity} summarizes the time and memory complexity of all algorithms introduced in this work.

\begin{table}[h]
\centering
\caption{Computational complexity of the algorithms introduced in this work.
$N$ is the number of qudits, $d$ the local Hilbert-space dimension ($d=2$ for qubits, $d=3$ for qutrits), $L$ the number of thermodynamic-integration grid points, and $n_S$ the number of Monte Carlo samples.}
\label{tab:complexity}
\begin{tabular}{@{}l@{\hspace{1.2em}}l@{\hspace{1.2em}}l@{\hspace{1.2em}}l@{\hspace{1.2em}}l@{}}
\toprule
Algorithm & Input & Time & Memory & Target \\
\midrule
Alg.~\ref{alg:sregray} (naive SRE) & $\ket{\psi}\in\mathbb{C}^{2^N}$ & $O(8^N)$ & $O(2^N)$ & SRE, exact \\
Alg.~\ref{alg:sre_eff} (FWHT SRE) & $\ket{\psi}\in\mathbb{C}^{2^N}$ & $O(N\cdot 4^N)$ & $O(2^N)$ & SRE, exact \\
Alg.~\ref{alg:sre_sampling_fht} (MC SRE) & $\ket{\psi}\in\mathbb{C}^{2^N}$ & $O(L\,n_S\,N\cdot 2^N)$ & $O(2^N)$ & SRE, sampling \\
Alg.~\ref{alg:mana_gray} (naive mana) & $\ket{\psi}\in\mathbb{C}^{3^N}$ & $O(27^N)$ & $O(3^N)$ & mana, pure, exact \\
Alg.~\ref{alg:mana_fht} (fast-FT mana) & $\ket{\psi}\in\mathbb{C}^{3^N}$ & $O(N\cdot 9^N)$ & $O(3^N)$ & mana, pure, exact \\
Alg.~\ref{alg:mana_mixed_fast_detailed} (mixed mana) & $\rho\in\mathbb{C}^{3^N\times 3^N}$ & $O(N\cdot 9^N)$ & $O(9^N)$ & mana, mixed, exact \\
\bottomrule
\end{tabular}
\end{table}

\subsection{Discussion and outlook}

Our algorithms are close to
optimal from a computational-complexity perspective.
Both for qubits (SRE) and for qutrits (mana), the efficient pure-state
algorithms evaluate all \(d^{2N}\) expectation values of Pauli strings or
phase-space point operators, respectively.
Given the overall cost \(O(N d^{2N})\), this implies that a \emph{single}
expectation value is obtained at a cost that scales only linearly with the
system size \(N\).
Among methods that explicitly compute the full generalized Pauli spectrum~\cite{Turkeshi25spectrum}
(i.e.\ all \(d^{2N}\) expectation values of generalized Pauli strings), our algorithms are therefore optimal up to a benign linear factor in \(N\).
Moreover, they operate at a memory cost proportional to that needed to
store the state vector of the underlying quantum state, without introducing any significant  additional memory overheads.

An interesting open question is whether one can bypass the explicit
evaluation of the full generalized Pauli spectrum and still obtain
numerically exact values of non-stabilizerness measures such as \(M_2\) or mana, with strictly lower asymptotic cost than \(O(N d^{2N})\).
Our results do not rule out such schemes, but they set a
natural benchmark for any algorithm that proceeds via expectation values.
A similar near-optimal behavior appears in the sampling setting: the
thermodynamic-integration scheme for SRE achieves an effective linear in
\(N\) computational cost per sampled Pauli string, and the mixed-state mana
algorithm requires only a linear-in-\(N\) overhead to evaluate a single
phase-space expectation value.
In this sense, both the exact and sampling-based approaches presented here
saturate, up to polynomial factors, the natural scaling dictated by the
number of independent expectation values that must be accessed to evaluate SRE and mana.

It is instructive to compare our approach with existing methods for
estimating non-stabilizerness.
The sampling algorithm of Ref.~\cite{Tarabunga25mutual} provides an
unbiased estimator of SRE by sampling Eq.~\eqref{eq:distNAT} at a cost scaling as
\(O(8^{N/2})\) per sample, without incurring any significant memory overhead.
However, in that scheme the variance of the Monte Carlo estimator for the SRE $M_q$ at $q>1$ can grow
exponentially with system size, so that maintaining a fixed statistical
precision may still require an exponential increase of the number of samples with $N$.
Another important class of approaches relies on tensor-network
representations of quantum states.
In Ref.~\cite{haug2023quantifying}, SRE is computed directly from a matrix
product state (MPS) approximation, with a cost scaling as \(\chi^{6q}\) for
integer R\'enyi index \(q\), where \(\chi\) is the MPS bond dimension.
Ref.~\cite{lami2023perfect} develops an MPS-based
sampler that draws \emph{independent} samples from the distribution
\(\pi(P)\) in Eq.~\eqref{eq:distNAT} at cost \(O(N\chi^3)\) per sample.
In Ref.~\cite{tarabunga2024mps}, SRE is
evaluated via contring MPS with a specifically designed matrix-product operator, with
a cost, after MPS compression, proportional to \(\chi^4\).
All these tensor-network-based methods are extremely powerful whenever the
entanglement of the state is modest, so that a relatively small bond
dimension \(\chi\) provides an accurate approximation and the cost remains
polynomial in \(N\).

In contrast to tensor-network approaches, the algorithms introduced in this
work operate directly on state-vector representations and make no
assumption of low entanglement, complementing the tensor network approaches.
The \href{https://github.com/bsc-quantic/HadaMAG.jl/}{HadaMAG.jl}
package is therefore particularly well suited for scenarios in which
many-body wave functions are explicitly available as state vectors and computing measures of non-stabilizerness via our exact state-vector
algorithms is the optimal strategy.
Prominent applications include non-equilibrium dynamics of quantum
circuits, where magic growth and spreading have been analyzed in detail
(e.g.~\cite{turkeshi2024magic,haug2024saturation,Aditya25spreading}), as
well as unitary dynamics of interacting many-body systems governed by
ergodic Hamiltonians or Floquet operators~\cite{Tirrito24anti,Tirrito25transp,Odavic25}.
Related settings encompass Mpemba effects~\cite{Aditya25mpemba} and local spreading of
magic~\cite{Aditya25spreading}. In these setups, the time-evolved states for comparatively large systems can be
generated efficiently by exploiting the structure of time evolution operator or the sparsity of the Hamiltonian matrix~\cite{Tal‐Ezer84}. Similarly, methods utilizing sparse Hamiltonian structure~\cite{Sierant20polfed} provide access to for highly excited eigenstates
 of many-body systems.
The same tools are directly applicable to non-ergodic phases such as
many-body localized systems~\cite{Sierant25mbl}, where SRE and mana have
already been used to characterize eigenstate properties and dynamics~\cite{Falcao25mbl}.
Other directions for applications involve constrained
dynamics and weak ergodicity breaking, including quantum many-body
scars~\cite{Serbyn21,Smith25scars}, as well as lattice gauge theories~\cite{Falcao25}.
Beyond these, \href{https://github.com/bsc-quantic/HadaMAG.jl/}{HadaMAG.jl} can be employed to probe magic in highly
excited mid-spectrum states~\cite{Turkeshi25spectrum} or in ground and
excited states of models with non-local interactions, such as
Sachdev--Ye--Kitaev-type systems and their variants~\cite{Bera2025SYK,Santra25complexity,Jasser2025}.
Finally, the algorithms are also well adapted to monitored and hybrid
unitary-measurement dynamics, where entanglement and magic exhibit rich
non-equilibrium phenomenology~\cite{fux2023entanglementmagic,bejan2023dynamical,Tarabunga2025transition, Paviglianiti25, Scocco25rise,Loio25telep}.

The combination of exact and sampling-based tools developed here paves the way for systematic, large-scale studies of magic dynamics and resource distribution across a broad range of many-body platforms. In all these settings, non-stabilizerness plays a central role in both enabling and constraining quantum advantage. We expect that the algorithms and software introduced in this work will serve as a practical baseline for future numerical investigations of non-stabilizerness, particularly in strongly entangled regimes where state-vector methods are the natural tools of choice.
Our results open also several conceptual avenues for exploration. The fast-transform perspective developed here suggests possibilities of generalizations to higher local dimensions and other non-stabilizerness measures, such as the generalized stabilizer entropies~\cite{turkeshi2024magic}, and may also
extend to other magic monotones and witnesses,
including systems with Clifford structure restricted, e.g., by symmetry \cite{Cepollaro25subspaces}).
In parallel, another open question is whether the fast-transform to evaluation of non-stabilizerness measures  can be utilized in experiments, e.g., by integrating it with
classical post-processing of randomized measurements or shadow
tomography. We leave these questions for future work.

\section{Acknowledgments} 
P.S. acknowledges insightful discussions with members of Quantic group at BSC and collaborations on related projects with S. Aditya, P.R.N. Falcão, J. Zakrzewski, E. Tirrito, X. Turkeshi.
P.S. acknowledges fellowship within the “Generación D” initiative, Red.es, Ministerio para la Transformación Digital y de la Función Pública, for talent attraction (C005/24-ED CV1), funded by the European Union NextGenerationEU funds, through PRTR. We acknowledge financial support from the Spanish Ministry for Digital Transformation and of Civil Service of the Spanish Government through the QUANTUM ENIA project call - Quantum Spain, EU through the Recovery,
Transformation and Resilience Plan – NextGenerationEU within the framework of the Digital Spain 2026.

{\it Note Added:} While finalizing this manuscript, we became aware of two independent works, Refs.~\cite{xiao2026exponentiallyacceleratedsamplingpauli, 
huang2026fastexactapproachstabilizer}, which employ fast Hadamard
transform to compute stabilizer R\'enyi entropies.
In parallel, as the \href{https://github.com/bsc-quantic/HadaMAG.jl/}{HadaMAG.jl}
codes were being developed and optimized, one of the authors (P.S.) applied the methods introduced here to the study of
non-stabilizerness in quantum circuits~\cite{turkeshi2024magic,
Aditya25mpemba,Aditya25spreading}, as well as in ergodic~\cite{Tirrito24anti} and non-ergodic~\cite{Falcao25mbl} quantum many-body systems.

\appendix

\section{Relation to FWHT-based approaches in prior work}
\label{app:fwht_comparison}

Several prior works have used fast Walsh--Hadamard or similar transforms in non-stabilizerness computations; here we compare them with the present algorithms, focusing on input representation and memory cost.

\paragraph{Ref.~\cite{Hamaguchi24handbook}.}
An FWHT-based overlap routine (Sec.~3) and an FWHT-like tensor-product transform (Appendix~B) serve two distinct purposes.
\textit{Appendix~B} computes all $4^N$ Pauli expectation values from the full density matrix via a tensor-product transform on the $4^N$-dimensional Choi vector: $O(N\cdot 4^N)$ time, but $O(4^N)$ memory---infeasible for state-vector inputs at $N=25$ ($4^{25}\approx10^{15}$ entries).
\textit{Section~3 (Theorem~1)} evaluates stabilizer overlaps $\mathbf{A}_n^\top\mathbf{b}$ for the robustness-of-magic linear program: for each of the $2^{O(n^2)}$ stabilizer groups, $2^N$ entries are extracted from the Pauli vector $\mathbf{b}$ and a FWHT on a $2^N$-dimensional vector is applied, with $O(2^N)$ extra working space. The routine requires the full $4^N$-component Pauli vector as input.

\paragraph{Ref.~\cite{Tarabunga25bmsa}.}
The basis-minimized stabilizerness asymmetry (BMSA) is computed using an analogous structure to \textit{Sec.~3} of Ref.~\cite{Hamaguchi24handbook}: a FWHT over $2^N$-dimensional subvectors of the Pauli vector, minimized over all $2^{O(n^2)}$ stabilizer groups. The routine requires enumerating all $2^{O(N^2)}$ stabilizer groups, limiting exact calculations to $N\leq 5$ qubits.

\paragraph{Pauli decomposition of operator matrices~\cite{Hantzko2024Tensorize, Jones2024Decomposing, VidalRomero2023}.}
These works compute all $4^N$ Pauli coefficients of a general $2^N\times 2^N$ operator matrix, motivated by Hamiltonian simulation and quantum chemistry.
Ref.~\cite{Hantzko2024Tensorize} achieves $O(N\cdot 4^N)$ time via tensorized recursive matrix slicing; Ref.~\cite{Jones2024Decomposing} achieves $O(8^N)$ time with $O(1)$ additional memory using Gray codes; Ref.~\cite{VidalRomero2023} achieves $O(8^N)$ time with $O(2^N)$ working memory via sparse Pauli products.
All three require the full $2^N\times 2^N$ matrix as input ($O(4^N)$ memory), and none is designed specifically for direct evaluation of the SRE or related magic monotones from a state vector.

\paragraph{The present work.}
All prior FWHT-based approaches discussed above are restricted to qubits ($d=2$) and require either the full density matrix or the $4^N$-entry Pauli vector as input. Algorithms~\ref{alg:sre_eff} and~\ref{alg:mana_fht} compute the SRE and mana directly from the $d^N$-dimensional state vector, achieving $O(N\, d^{2N})$ time and $O(d^N)$ memory without constructing a density matrix or Pauli vector. In particular, the mana algorithms extend the fast-transform approach to qutrits ($d=3$). Algorithm~\ref{alg:mana_mixed_fast_detailed} takes the density matrix as input ($O(9^N)$ memory) but applies the transform in place. Algorithm~\ref{alg:sre_sampling_fht} samples over $X$-patterns and evaluates each $Z$-sum via a single fast transform, maintaining $O(2^N)$ memory throughout.

%\bibliography{bib}

\end{document}